\documentclass[11pt,a4paper]{article}
\usepackage{jheppub}
\usepackage[T1]{fontenc}
\usepackage[utf8]{inputenc}
\usepackage{dcolumn}
\usepackage{bm,amsfonts,amsthm,amsmath,amssymb}

\usepackage{mathrsfs}
\usepackage{mathtools}
\usepackage{comment}
\usepackage{color}
\usepackage{enumerate}
\usepackage{subfigure}

\DeclareMathOperator{\sech}{sech}
\DeclareMathOperator{\Tr}{Tr}

\newcommand{\bi}{\begin{itemize}}
\newcommand{\ei}{\end{itemize}}
\newcommand{\grad}{\ensuremath{\vec{\nabla}}} 
\newcommand{\be}{\begin{equation}}
\newcommand{\ee}{\end{equation}}
\newcommand{\nn}{\nonumber}
\newcommand{\CC}{\ensuremath{\mathbb{C}}}
\newcommand{\ZZ}{\ensuremath{\mathbb{Z}}}
\newcommand{\RR}{\ensuremath{\mathbb{R}}}
\newcommand{\aut}{\ensuremath{{\rm aut}}}
\newcommand{\mff}{\ensuremath{\mathfrak{f}}}
\newcommand{\mfg}{\ensuremath{\mathfrak{g}}}
\newcommand{\Vcal}{\ensuremath{{\cal V}}}
\newcommand{\Ocal}{\ensuremath{{\cal O}}}
\newcommand{\Ccal}{\ensuremath{{\cal C}}}
\newcommand{\Hcal}{\ensuremath{{\cal H}}}
\newcommand{\Ical}{\ensuremath{{\cal I}}}
\newcommand{\Pcal}{\ensuremath{{\cal P}}}

\newcommand{\rhs}{\ensuremath{{\rm RHS}}}

\newcommand{\Ncal}{\ensuremath{{\cal N}}}

\newcommand{\ndim}{d}
\renewcommand{\Re}{{\rm Re}}

\title{Unitarity at the Late time Boundary of de Sitter}
\author[a]{Gizem \c{S}eng\"or}
\author[a]{Constantinos Skordis}
\affiliation[a]{CEICO, Institute of Physics of the Czech Academy of Sciences,\\ Na Slovance 1999/2, 182 21 Praha 8, Czechia}
\emailAdd{sengor@fzu.cz}
\emailAdd{skordis@fzu.cz}
\abstract{
The symmetry group of the de Sitter spacetime, accommodates fields of various masses and spin among its unitary irreducible representations. 
These unitary representations are labeled by the spin and the weight contribution to the scaling dimension
and depending on the mass and spin of the field the weight may take either purely real or purely imaginary values. 
In this work, we construct the late time boundary operators for a massive scalar field propagating in de Sitter spacetime, in arbitrary dimensions. 
  We show that contrary to the case of Anti de Sitter, purely imaginery weights also correspond to unitary operators, as well as the ones with real weight, 
 and identify the corresponding unitary representations.
 We demonstrate that these operators correspond to the late time boundary operators 
 and elucidate that all of them have positive definite norm.}
\arxivnumber{1912.09885}

\begin{document}
\maketitle

\section{Introduction}
The de Sitter spacetime~\cite{deSitter:1917zz} is one of the maximally symmetric solutions to the vacuum Einstein equations in the presence of a positive cosmological constant.
Over the last forty years or so  it has acquired particular significance in cosmology. 
The latest cosmological observations~\cite{Beutler:2011hx,Ross:2014qpa,Betoule:2014frx,Bautista:2017wwp,Scolnic:2017caz,Aghanim:2018eyx,Abbott:2018wzc,Akrami:2018odb} 
indicate that our Universe is well described  by the inflationary $\Lambda$CDM model.  
While several tensions between various data sets have been noted~\cite{Riess:2018uxu,Riess:2019cxk,Colin:2018ghy,Joudaki:2016kym,Motloch:2018pjy,Ade:2013lmv,Sakr:2018new}, 
the  $\Lambda$CDM  model remains the simplest model that can accommodate the majority of observations. Taken at face value, this model points to two eras 
where the de Sitter spacetime is relevant: the past de Sitter phase which may be 
considered as a limit of most inflationary models and future de Sitter phase that the universe will eventually enter, that is the era of dark energy.


The de Sitter (dS) manifold describes a spacetime whose 
spatially flat sections expand in an accelerated fashion with time. In terms of global coordinates the spatial hypersurfaces are $3$-spheres that grow with time. 
Hence, the de Sitter manifold sets a good background for the study of inflationary and dark energy eras.
For instance the propelling observable quantities in inflationary studies are inflationary correlators
and the characteristics of these correlators can give insight into the type of fields and their interactions present during inflation. 
It is possible to catalog expected forms for these correlators in the presence of heavy mediator particles by considering the restrictions 
due to the conformal symmetries the de Sitter background posses \cite{Creminelli:2011mw}. See \cite{Baumann:2019oyu} for a recent example that elaborates 
this discussion to the level of a cosmological bootstrap mechanism by focusing on the four-point function of conformally coupled and massless scalars in de Sitter where interactions are mediated by exchange of massive and non-zero spin particles. 

The superhorizon scale and correspondingly the late time limit behavior of de Sitter correlators is of particular interest.
 Recently, in \cite{Mirbabayi:2019qtx} the behavior of the scalar perturbation correlator on superhorizon scales has been recast
 in the static patch of de Sitter in order to investigate  how long it takes for a perturbed state,  considered as an excited state, 
to relax back to the equilibrium state defined as the vacuum wavefunction in global de Sitter. 
It is shown that in the late time limit, this process is a Markovian evolution, which means the evolution does not depend on the history of the process for slowroll potentials.

  An analysis of quantized fields on de Sitter, on one hand involves field equations derived from an action principle, the vacuum state, the Hilbert space 
and the Green functions \cite{Chernikov:1968zm, Bunch:1978yq,Sasaki_1995}. On the other hand one can start by considering the unitary irreducible representations of the 
symmetry group associated with de Sitter and construct local fields from these representations. To list a few examples from earlier work where group theoretical approaches 
have been eluminating, the uniqueness of the de Sitter invariant vacuum state have been recognized to be less ambiguous in a group theoretical approach 
in two-dimensional de Sitter \cite{Joung:2006gj,Joung:2007je}. Superradiance has been restudied with a group theoretical approach to near horizon geometry 
of charged rotating black holes for gaining more insight \cite{Anninos:2019oka}. In these works the 
principal series representations of two-dimensional de Sitter and two-dimensional Anti de Sitter (AdS) respectively, 
play an important role. In the context of modified gravity, the tensor perturbations have been identified to belong to the discrete series representations of 
four-dimensional de Sitter \cite{Dehghani_2016}. Last but not least is an example from Mean Field Theory, where the operator product expansion coefficients are obtained 
via group theory methods \cite{Karateev:2018oml}. 

In this work, our main concern is the relation between the irreducible representations of the de Sitter symmetry group and the late time behavior of massive scalar fields $\phi$
of mass $m$ on de Sitter.  Our aim is to demonstrate that all of the representations involved are unitary.

 The Killing vectors of de Sitter generate symmetries each of which is a subgroup of the group $SO(2h+1,1)$ where $h$ is a half-integer. Therefore the symmetry group of $2h+1$-dimensional 
de Sitter, from now on denoted as $dS_{2h+1}$, is the group $SO(2h+1,1)$. This group also happens to be the conformal group of $2h$ dimensional Euclidean space. 
Considering the metric, the connection between the $2h$ dimensional Euclidean space and $dS_{2h+1}$ becomes explicit under an early or late time limit. Historically, 
the representation theory of $SO(2h+1,1)$ has been heavily studied by Harish-Chandra \cite{harish-chandra1970}. The unitary irreducible representations
 of $SO(2h+1,1)$ are labeled and categorized by their spin $\ell$ and \emph{weight} $c$ of their \emph{scaling dimension} $\Delta = h + c$. The categories are referred to as \emph{principal series, 
complementary series, exceptional series} and \emph{discrete series}. In this work we focus on the principal and complementary series.

The fact that the scaling dimension is decomposed into $\Delta = h + c$ is not a coincidence. Indeed, for the group $O(d+1,1)$ the 
quantity $\frac{d}{2}$ is the half sum of the restricted positive roots. 
In the convention of \cite{Dobrev:1977qv} it is denoted by $\frac{d}{2}\equiv h$ and the scaling dimension for any spin is decomposed in this way. 
Here, we also work with the notation $h$ and $c$ in order to emphasize the role of the scaling weight $c$ in the categorization of unitary irreducible representations. 
Depending on the mass $m$ of the field, the weight $c$ can be either purely real or purely imaginery. 
 To put it more explicitly, letting $H$ be the Hubble constant of de Sitter, light fields with mass $\frac{m^2}{H^2}< h^2$ correspond 
to real weight $c$ representations while heavy fields whose mass are in the range $\frac{m^2}{H^2} > h^2$, have imaginary weight $c$ (and hence complex scaling dimension $\Delta$).


For comparison, the manner in which the unitary irreducible representations of the other maximally symmetric spacetimes are classified is different than that of de Sitter spacetime. 
In the case of Anti de Sitter spacetime where the relevant group is $SO(2h,2)$, imaginary scaling weight does not arise among unitary irreducible representations 
and therefore there is no distinction between light and heavy fields \cite{Basile:2016aen}\footnote{In the reference \cite{Anninos:2019oka} principal series representations of $AdS_2$ arise because of the charge of the field.}. In contrast,
 we discuss below that light fields on de Sitter belong to the complementary series representations while heavy fields belong to principal series representations. 
The work~\cite{Basile:2016aen}, which is complementary to ours,
 is a good reference for readers interested in comparing the unitary representations of Anti de Sitter with those of de Sitter. 
We note that their aims are very different than ours and thus they follow a different approach.
 There, the authors  studied the behavior of fields of any spin on (A)dS
 under a specific limit in which (A)dS spacetimes approach Minkowski by focussing on the characters of unitary irreducible representations of the de Sitter group.
Moreover, \cite{Basile:2016aen} demand that the fields involved are well-behaved at the late time boundary and this fixes their boundary conditions. 

In the case of Minkowski spacetime, as nicely reviewed in \cite{bekaert2006unitary}, the induced unitary irreducible representations 
of the Poincaré group, $ISO(2h,1)$, are classified into \emph{massive},\emph{massless}, \emph{tachyonic} and \emph{zero momentum representations}. 
The massless representations divide further into two categories of \emph{helicity representations} and \emph{infinite (continuous spin) representations}.
 The tachyonic representations are unitary, however, they are not causal and hence they are not considered physical. 
The zero momentum and the tachyonic representations in $2h+1$ dimensions correspond to the unitary representations of one lower dimensional de Sitter space $dS_{2h}$
and two lower dimensional de Sitter $dS_{2h-1}$ respectively.

There also exist limits between certain de Sitter representations and Poincaré representations. 
For instance \cite{mickelsson1972} show the İnönü-Wigner contraction of the principles series representations 
of de Sitter to positive mass representations of Poincare. Also see \cite{unknown} and references 
there in for a summary of further discussion on limits between the representations of  de Sitter 
and Poincaré and Anti de Sitter and Poincaré groups. For more geometrically oriented readers, we also note the reference \cite{huguet2016massive} where they derive 
the Klein-Gordon equation for massive classical fields on de Sitter and Anti-de Sitter 
from a geometric point of view. They point out that as a consequence of geometry the scalar field 
is a homogeneous function where the scaling dimension is also the degree of homogeneity.

This work is organised as follows. In section \ref{sec:ltandboundaryops} we consider massive scalar fields $\phi$ of mass $m$ propagating in de Sitter spacetime and 
determine their behaviour in the late time limit. It is well known that in the late time limit the field takes the form of $\phi(\vec{x},\eta) \rightarrow  \eta^\Delta \Ocal_\Delta(\vec{x})$
and our aim in that section is to explicitly identify the operators $\Ocal_\Delta(\vec{x})$. We carefully take the late time limit of the bulk solution 
and identify $\Ocal_\Delta$ in momentum space, in terms of creation and annihilation operators $a^\dagger_{\vec{q}}$ and $a_{\vec{q}}$ of state with momentum $\vec{q}$.
As we present in section \ref{sec:ltandboundaryops}, for such a  scalar field solution, $\Delta$ depends on the mass of the field and dimensions 
	of spacetime (see ~\cite{Monten:2017urt} for the transformation properties of $\Ocal_\Delta$ under $SO(2h+1,1)$). These operators are the first result of the paper, and are summarized in table \eqref{summary of the operators}. 

Before demonstrating the unitarity of the boundary operators, we follow the indepth monograph \cite{Dobrev:1977qv} 
on the group $SO(2h+1,1)$ in section \ref{sec:elementsofG}, 
in order to review the group properties and irreducible representations. This technical section also works to introduce the notation. 
In short, the unitary irreducible representations are denoted as $\chi=\{\ell,c\}$ and are 
realized by functions $f(\vec{x})$, whose properties are specified in section \ref{sec:connection to functions over coordinate space}. 
These functions are in correspondence with the  boundary operators $\Ocal_\Delta$.

In section \ref{sec:unitaryreps} we use the inner product of  \cite{Dobrev:1977qv} applied to 
the irreducible representations of de Sitter and review the conditions on the weights for these representations to be unitary. 
Contrary to intuition the inner product for the operators with imaginery weight is less subtle than that for operators with real weight.
 The definition of a unitary inner product for representations with real weight involves a so-called intertwining operator. 

The subtleties of the inner product are presented for both real and imaginary weights in section \ref{sec:positivenormandunitarity}, 
by making use of the boundary operators identified in section \ref{sec:ltandboundaryops}. There, we demonstrate 
the unitarity of the boundary operators, which is the main goal of the paper. The boundary operators we obtain are given in momentum space but the form $\Delta=h+c$ for the scaling dimension holds in position space. 
In section \ref{sec:connection between intertwiner and shadow tr} we confirm that our operators indeed have the expected scaling dimension and demonstrate how the intertwining operator leads to a shadow transformation.
We conclude in section~\ref{sec:conclusion} where we give a summary of our results, have a short discussion regarding the notion
of unitarity on de Sitter and consider  possible future directions.

In appendix \ref{app:furtherpropertiesofG} we present some properties of the de Sitter group $SO(2h+1,1)$, in appendix \ref{app:normGX2} we present details regarding
 the normalization of the intertwining operator from \cite{Dobrev:1977qv} and in appendix \ref{app:exemplary shadow transforms} we give a concrete example of a shadow
transformation.

We use a mostly positive signature convention and throughout this article we use the convention that Greek indices take values in the range $0 \ldots 2h+1$,
uppercase Latin indices in the range $1 \ldots 2h+1$ and lowercase Latin  in the range $1 \ldots 2h$.

	 \section{Late time behavior and boundary operators}
	 \label{sec:ltandboundaryops}
	 \subsection{Scalar Fields}
There is a large variety of coordinate systems one can use to describe de Sitter (see \cite{Anninos:2012qw}), each of which has its own merits
or shortcomings. For our purposes, we use the planar coordinates, covering only half of the de Sitter manifold, for which the de Sitter metric takes the form
 \be 
  ds^2=\frac{-d\eta^2+d\vec{x}^2}{H^2\eta^2},
\ee
where $\eta$ is the conformal time.  In these coordinates the action of a  massive scalar field propagating on a $2h+1$ dimensional de Sitter manifold,
takes the form
	 \begin{align}
	 S&=\int d^{2h+1}x\sqrt{-g}\left[-\frac{1}{2}g^{\mu\nu}\partial_\mu\phi\partial_\nu\phi-\frac{1}{2}m^2\phi^2\right]\\
	 &=\frac{1}{2}\int \frac{d^{2h+1}x}{\left(H\eta\right)^{2h+1}}\left\{
  H^2\eta^2\left[\phi'^2-\left(\grad\phi\right)^2\right] 
 -m^2\phi^2\right\},
	 \end{align}
where a prime denotes differentiation wrt $\eta$, and leads to the following equation of motion
\be 
\phi''-\frac{2h-1}{\eta}\phi'-\grad^2\phi+\frac{m^2}{H^2\eta^2}\phi=0.
\ee
	 It is more convenient to expand the field $\phi(\vec{x},\eta)$ in terms of its Fourier modes as follows
	 \be 
 \phi(\vec{x},\eta)=\int\frac{d^{2h}q}{(2\pi)^{2h}}\left[\phi_q(\eta)a_{\vec{q}}\, e^{i\vec{q}\cdot\vec{x}} + \phi^*_q(\eta)a^\dagger_{\vec{q}} \, e^{-i\vec{q}\cdot\vec{x}}\right],
\ee
	 where in quantizing the field the coefficients $a_{\vec{q}}$ and $a^\dagger_{\vec{q}}$ obey
	 \be 
\left[a_{\vec{q}},a^\dagger_{\vec{q}'}\right]=(2\pi)^{2h}\delta^{2h}(\vec{q}-\vec{q}\,').
\ee
	 The equation of motion for the mode functions $\phi_q(\eta)$ then reads
	 \be\label{eommode} \phi_q''-\frac{2h-1}{\eta}\phi'_q+\left(q^2+\frac{m^2}{H^2\eta^2}\right)\phi_q=0.\ee
	 Equation \eqref{eommode} is in fact Bessel's equation in disguise and this becomes more apparent if 
 one defines $\phi_q(\eta)\equiv \eta^{h}\varphi(q\eta)$ and $u\equiv q\eta$. 
  Then equation \eqref{eommode} turns into
\be 
  \frac{d^2}{du^2}\varphi(u)+\frac{1}{u}\frac{d}{du}\varphi(u)+\left[1+\left(\frac{m^2}{H^2}-h^2\right)\frac{1}{u^2}\right] \varphi(u) =0,
\label{Bessel_eq}
\ee
	whose solutions are the Bessel functions of the first and second kind, $J_\nu(q\eta)$ and $Y_\nu(q\eta)$ with 
\begin{equation}
 \nu^2=h^2-\frac{m^2}{H^2}.
\label{nu_h_m}
\end{equation}
	 The solution for the mode functions $\phi_q(\eta)$ that approaches the Bunch-Davies vacuum solution at early times $|\eta|\to\infty$ is
\be 
  \phi_q(\eta)=|\eta|^{h}H^{(1)}_\nu(q|\eta|)
\ee
	with $H^{(1)}(q|\eta|)=J_\nu(q|\eta|)+iY_\nu(q|\eta|)$, the Hankel function of the first kind. 
	Thus the full solution for the scalar field in the bulk is
	\be 
 \label{phibulksoln} 
\phi(\vec{x},\eta)=\int\frac{d^{2h}q}{(2\pi)^{2h}}\left[|\eta|^{h}H^{(1)}_\nu(q|\eta|)a_{\vec{q}}e^{i\vec{q}\cdot\vec{x}}+|\eta|^{h}H^{*(1)}_\nu(q|\eta|)a^\dagger_{\vec{q}}e^{-i\vec{q}\cdot\vec{x}}\right]
\ee
	which is normalized\footnote{From dimensional analysis one can see that the normalization of $\phi(\vec{x},\eta)$ will involve $H^{h-1/2}$. In the rest of the manuscript we drop this factor of $H$ and any other numerical factors such as $\pi$ in the overall normalization.}  with respect to the Klein-Gordon inner product \cite{Birrell:1982ix}.
	
%
	 At late times as $q\eta\to 0$, the spatial dependence in equation \eqref{eommode} is negligible, and the late time solution approaches the following form
\be
\label{ltformat}
\lim_{\eta\to 0}\phi(\vec{x},\eta)=\Ocal_\Delta(x)\eta^{\Delta}~~\text{where}~~\Delta=h\pm\nu.
\ee
	 On one hand, it is the term $\Ocal_\Delta(x)$ that can be recognized as an operator of dimension $\Delta$ at the late time boundary of de Sitter and there is expected to be a corresponding CFT current of dimension $2h-\Delta$ in the dual  Euclidean CFT. On the other hand $\Ocal_\Delta(x)$ is an irreducible representation of $SO(2h+1,1)$. 
We review the properties of irreducible representations in section \ref{sec:elementsofG} and demonstrate how they are categorized in section \ref{sec:unitaryreps}. In this section we show how to read-off $\Ocal_\Delta(x)$ for different mass ranges.
	 
	 By definition, the scaling dimension $\Delta$ is decomposed into two parts as follows
	 \be \label{deltaformat} \Delta=h+c.\ee
	 This decomposition emphasizes the contribution of the dimensionality of spacetime by $h$ and the contribution of the properties of the field, namely spin and mass, by the weight $c$.  

	 Comparing \eqref{ltformat} with \eqref{deltaformat} we see that
	 \be c\equiv \nu.\ee
	 In the coming sections we will see that $c$ being real or purely imaginery  plays an important role in the treatment of the corresponding boundary operators. As a first step let us determine what the late time boundary operators are. 


The decomposition of the scaling dimension into these two specific parts has to do with the properties of $O(d+1,1)$. For the group $O(d+1,1)$, $h$ is the half sum of the restricted positive roots \cite{Dobrev:1977qv}. The parameter $c$, which will be referred to as \emph{the scaling weight} from now on, 
is in general related to both the mass and spin of the field under consideration. The irreducible representations are labeled by their spin $l$, and the 
weight $c$ of the scaling dimension $\Delta=h+c$. We will work with the notation $h$ and $c$ to emphasize the role of the scaling weight $c$ in the 
categorization of unitary irreducible representations. 
	 
	 \subsection{Light scalars with mass $\frac{m^2}{H^2}=\frac{(2h)^2-1}{4}$}
	 \label{subsection:special mass soln}
	 Conformally coupled scalar fields on $dS_3$ and $dS_4$ have masses $\frac{m^2}{H^2}|_{dS_3}=\frac{3}{4}$ and $\frac{m^2}{H^2}|_{dS_4}=2$ respectively, which satisfy the condition 
	 \be \label{specialmasscasedefined} \frac{m^2}{H^2}=\frac{(2h)^2-1}{4}~~\text{where}~~\nu=\pm\frac{1}{2}.\ee
	 These fields also accommodate the scalar sector of Vasiliev Higher Spin gravity on de Sitter.
	 
	 In this case, the Hankel functions have a very simple form to work with
	 \be H^{(1)}_{\pm\frac{1}{2}}(q|\eta|)\simeq\frac{e^{-iq\eta}}{\sqrt{q|\eta}}.\ee
	 The only difference between $\nu=+\frac{1}{2}$ and $\nu=-\frac{1}{2}$ is an overall numerical coefficient which can be dismissed for our purposes. Thus we have the following mode functions
	 \be \label{specialsoln}
	 \phi_q(\eta)=\eta^{\Delta_{-}}\frac{e^{-iq\eta}}{\sqrt{2q}} \quad~\text{with}~\quad \Delta_{-}=h-\frac{1}{2}\quad~\text{for}~\frac{m^2}{H^2}=h^2-\frac{1}{4},\ee
	 in this case. 
	 
	 At this point we would like to read off the expression for the boundary operators $\Ocal_\Delta$ by matching the late time limit of the general solution
	 \be\label{bulkspecialcase}\phi(\vec{x},\eta)=\int \frac{d^{2h}q}{(2\pi)^{2h}}\frac{\eta^{\Delta_{-}}}{\sqrt{2q}}\left[a_{\vec{q}}e^{-iq\eta+i\vec{q}\cdot\vec{x}}+a^\dagger_{\vec{q}}e^{iq\eta-i\vec{q}\cdot\vec{x}}\right],\ee
	 to the expected form of \eqref{ltformat}. Following the procedure outlined in \cite{Anninos:2017eib} for the case of Vasiliev scalar on $dS_4$,
 rewriting \eqref{bulkspecialcase} as
	 \be\label{trickformat} \phi(\vec{x},\eta)=\int\frac{d^{2h}q}{(2\pi)^{2h}}\frac{\eta^\Delta}{\sqrt{2q}}\left[f_1(q)\cos(q\eta)+f_2(q) \sin(q\eta)\right]e^{i\vec{q}\cdot\vec{x}}\ee
	 makes it easier to take the late time limit. We can achieve this by sending $\vec{q}\to-\vec{q}$ in the second term of \eqref{bulkspecialcase}. In this limit $d^3q\to d^3q$, $\vec{q}\cdot\vec{x}\to-\vec{q}\cdot\vec{x}$, $q\to q$ and $a^\dagger_{\vec{q}}\to a^\dagger_{-\vec{q}}$, leading to
	 \be \phi(\vec{x},\eta)=\int\frac{d^{2h}q}{(2\pi)^{2h}}\frac{\eta^\Delta}{\sqrt{2q}}\left[a_{\vec{q}}e^{-iq\eta}+a^\dagger_{-\vec{q}}e^{iq\eta}\right]e^{i\vec{q}\cdot\vec{x}}.\ee
	 Lastly, expanding the exponential functions in terms of sines and cosines gives
	 \be \phi(\vec{x},\eta)=\int\frac{d^{2h}q}{(2\pi)^{2h}}\frac{\eta^{\Delta_{-}}}{\sqrt{2q}}\left[\left(a_{\vec{q}}+a^\dagger_{-\vec{q}}\right)\cos(q\eta)-i\left(a_{\vec{q}}-a^\dagger_{-\vec{q}}\right)\sin(q\eta)\right]e^{i\vec{q}\cdot\vec{x}},\ee
	 hence,
	 \begin{subequations}
	 	\label{thefs}
	 	\begin{align}
	 	f_1(q)=a_{\vec{q}}+a^\dagger_{-\vec{q}}\\
	 	f_2(q)=-i\left(a_{\vec{q}}-a^\dagger_{-\vec{q}}\right)
	 \end{align}
	 \end{subequations}
	 
	  The family of solutions with $\nu=\pm\frac{1}{2}$ are related to each other by $\Delta_+=\Delta_-+1=\Delta+1$. Since $\cos(q\eta)\to1$ and $\sin(q\eta)\to q\eta$ as $\eta\to0$, the format \eqref{trickformat} in the late time limit becomes
\begin{align}
\nn
\lim_{\eta\to0}\phi(\vec{x},\eta)
 &= \int \frac{d^{2h}q}{(2\pi)^{2h}}\left[\eta^{\Delta_-}\frac{f_1(q)}{\sqrt{2q}}+\eta^{\Delta_+}\sqrt{\frac{q}{2}}f_2(q)\right]e^{i\vec{q}\cdot\vec{x}}
\\
	  & \stackrel{!}{=}\int \frac{d^{2h}q}{(2\pi)^{2h}}\left[\eta^{\Delta_-}\alpha(\vec{q})+\eta^{\Delta_+}\beta(\vec{q})\right]e^{i\vec{q}\cdot\vec{x}}.
\label{LTL_form}
\end{align}
	  By equations \eqref{thefs} and \eqref{LTL_form} we can read off that for scalar fields whose mass satisfy the relation $\frac{m^2}{H^2}=h^2-\frac{1}{4}$, the boundary operators are 
	  \begin{subequations}
	  	\label{boundaryoperators}
	  	\begin{align}
	  	\alpha(\vec{q})&=\frac{a_{\vec{q}}+a^\dagger_{-\vec{q}}}{\sqrt{2q}}\\
	  	\beta(\vec{q})&=-i\sqrt{\frac{q}{2}}\left(a_{\vec{q}}-a^\dagger_{-\vec{q}}\right).
	  	\end{align}
	  \end{subequations}     
	 irrespective of the dimension $2h$. This matches the operators considered in \cite{Anninos:2017eib} up to an overall minus sign and a normalization convention. 
	 Hence forth, we use the notation that the lower weight operator is denoted by $\alpha$ and the higher weight by $\beta$.
	 
	 \subsection{Light Scalars in General}
	 \label{sec:light scalars in general}
	Notice that since
	 \be c\equiv \nu =\sqrt{h^2-\frac{m^2}{H^2}},\ee
	 the weight $c$ can be real or purely imaginary depending on the mass of the scalar. For light scalars with masses in the range 
	 \be \frac{m^2}{H^2} < h^2 ,\ee
	 the weight $c$ is a real number and the previous case of $\frac{m^2}{H^2}=h^2-\frac{1}{4}$ is in fact contained in this 
range as a special case. We now discuss light fields more generally. Depending on the value of $\nu$, i.e. whether it is positive or negative or integer, 
the late time limit of Bessel functions for obtaining the boundary operators needs to be taken with some care.
	 
	 When $\Re(\nu)>0$ or when $\nu=-\frac{1}{2},-\frac{3}{2},-\frac{5}{2}$ in the limit $q|\eta|\to0$ the Bessel functions behave as
	 \begin{subequations}
	 	\begin{align}
	 	\lim_{q|\eta|\to0}J_\nu(q|\eta|)&\to \frac{1}{\Gamma(\nu+1)}\left(\frac{q|\eta|}{2}\right)^\nu,\\
	 	\lim_{q|\eta|\to0}Y_\nu(q|\eta|)&\to-\frac{\Gamma(\nu)}{\pi}\left(\frac{q|\eta|}{2}\right)^{-\nu}.
	 	\end{align}
	 \end{subequations}
	 Comparing the asymptotic behaviors of $J_\nu(q|\eta|)$ and $Y_\nu(q|\eta|)$, as the argument tends to zero the branch with the negative exponent will dominate over the branch with the positive exponent. Hence, in practice $\lim\limits_{z\to 0}H_{\nu}^{(1)}(z)\sim z^{-|\nu|}$. But this practical expression will involve only one of the two boundary operators, the one with the lower scaling dimension $\Delta=h-|\nu|$. As we want to be knowledgeable of both operators 
 we take instead the limiting form as\footnote{In what follows, our main concern will be the $q$-dependence and the $a_{\vec{q}}$, $a^\dagger_{\vec{q}}$ structure in the late time behavior, rather than the numerical factors.}  
	 \begin{align}
	 	\lim_{q|\eta|\to0}H^{(1)}_\nu(q|\eta|)&\to\frac{1}{\Gamma(\nu+1)}\left(\frac{q|\eta|}{2}\right)^\nu-i\frac{\Gamma(\nu)}{\pi}\left(\frac{q|\eta|}{2}\right)^{-\nu}.
	 	 	\end{align}
	 
	 With this asymptotic behavior, the general solution \eqref{phibulksoln} 
 when $\nu>0$ or when $\nu=-\frac{1}{2},-\frac{3}{2},...$, in the late time limit turns into
\begin{align}
\label{lightlt}  
\nn
\lim\limits_{z\to0}\phi(\vec{x},\eta)=\int\frac{d^{2h}q}{(2\pi)^{2h}}&
 \Bigg\{ |\eta|^{h-\nu}\left[
 -i \frac{\Gamma(\nu)}{\pi}  \, \left(\frac{q}{2}\right)^{-\nu} \,
 \left(a_{\vec{q}}-a^\dagger_{-\vec{q}}\right)
\right]
\\
  &+ |\eta|^{h+\nu}\left[\frac{1}{\Gamma(\nu+1)} \, \left(\frac{q}{2}\right)^\nu \, \left( a_{\vec{q}}+a^\dagger_{-\vec{q}} \right)
\right]\Bigg\} e^{i\vec{q}\cdot\vec{x}}
\end{align}	  
There is some subtlety in reading off the boundary operators from \eqref{lightlt}, depending on what $\nu$ is being considered. Keeping in mind that so far our notation has been such that the operator with the bigger scaling dimension is denoted as $\beta$ and the one with the smaller dimension as $\alpha$, $\Delta_\beta>\Delta_\alpha$, we consider the late time solution \eqref{lightlt} in two branches.

\subsubsection{The branch when $\nu>0$}
When $\nu>0$ the exponents are ordered as 
\be h-\nu ~<~ h+\nu.\ee
Thus we identify the boundary operators as
\begin{subequations}
	\label{branchrevg0boundary}
\begin{align}
\alpha^{I}(\vec{q}) =&  - i \frac{\Gamma(\nu)}{\pi} \, \left(\frac{q}{2}\right)^{-\nu} \,
  \left(a_{\vec{q}}-a^\dagger_{-\vec{q}}\right)~~&\text{with}~~\Delta^{I}_\alpha=h-\nu,
\\
 \beta^{I}(\vec{q})=& \frac{1}{\Gamma(\nu+1)} \, \left(\frac{q}{2}\right)^{\nu} \, \left( a_{\vec{q}}+a^\dagger_{-\vec{q}}  \right) ~~&\text{with}~~\Delta^{I}_\beta=h+\nu.
\end{align}
\end{subequations}

\subsubsection{The branch when $\nu=-\frac{2n+1}{2}$}
In this branch $\nu=-\frac{1}{2},-\frac{3}{2},etc.$, and the exponents are ordered as
\begin{align}
h-\nu=h+\frac{2n+1}{2}~>~h+\nu=h-\frac{2n+1}{2}.\end{align}
Thus, for $\nu=-\frac{2n+1}{2}$ where $n=0,1,2,etc.$, the identification of the boundary operators are as 
\begin{subequations}
	\label{-2n+1/2branch}
\begin{align}
\alpha^{II}(\vec{q}) =&  \frac{1}{\Gamma(\frac{1-2n}{2})} \, \left(\frac{q}{2}\right)^{-\frac{2n+1}{2}} \left( a_{\vec{q}}+a^\dagger_{-\vec{q}}\right)~~&\text{with}~~\Delta^{II}_\alpha=h-\frac{2n+1}{2},
\\
\beta^{II}(\vec{q})=& 
  -i  \frac{\Gamma(-\frac{2n+1}{2})}{\pi}  \left(\frac{q}{2}\right)^{\frac{2n+1}{2}} \left(a_{\vec{q}}-a^\dagger_{-\vec{q}}\right)~~&\text{with}~~\Delta^{II}_\beta=h+\frac{2n+1}{2}.
\end{align}
\end{subequations} 

The special case of scalar fields with mass $\frac{m^2}{H^2}=h^2-\frac{1}{4}$ presented in section \ref{subsection:special mass soln} 
  belongs to this branch with $\nu=-\frac{1}{2}$. Indeed letting $n=0$, so that $\Gamma\left(-\frac{1}{2}\right)=-2\sqrt{\pi}$ 
  and $\Gamma\left(\frac{3}{2}\right)=\frac{\sqrt{\pi}}{2}$ equations \eqref{-2n+1/2branch} give
\begin{subequations}
\begin{align}
\alpha^{II}(\vec{q})=&\sqrt{\frac{2}{\pi q}}\left( a_{\vec{q}}+a^\dagger_{-\vec{q}}\right)
\\
\beta^{II}(\vec{q})=& i\sqrt{\frac{2q}{\pi}}\left(a_{\vec{q}}-a^\dagger_{-\vec{q}}\right).
\end{align}
\end{subequations} 
These expressions match the previous solutions presented in \eqref{boundaryoperators} for this case, up to a factor of two and a minus sign in the case of $\beta^{II}$. They also match the solutions of \cite{Anninos:2017eib}, again up to numerical factors.

\subsubsection{The connection between the two branches}
\begin{figure}
 	   \center{\includegraphics[width=0.7	 \textwidth]{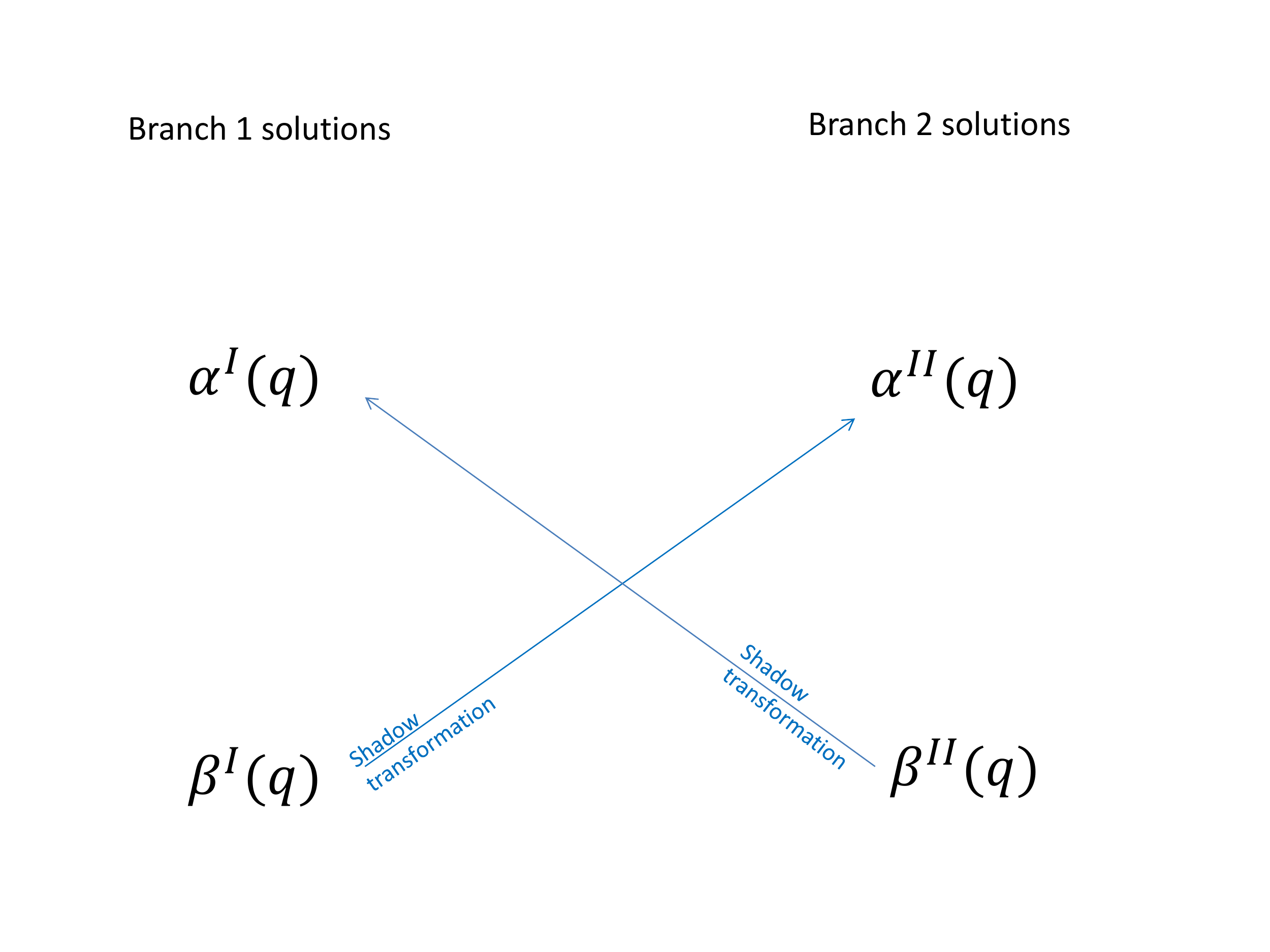}}
 	  \caption{\label{fig:shadowbetweenbranches} The shadow transformation takes the branch 2 light field solutions to the branch 1 solutions.}
 	\end{figure}

We have so far split the light scalar operators into two branches in order to better classify the lower and higher dimensional operators. 
 Among the first branch operators, the ones that arise for $\nu=\frac{2n+1}{2}$ will have scaling dimensions $h\pm\frac{2n+1}{2}$. 
  These scaling dimensions also come up in the operators of the second branch, suggesting a possible connection between the 
 two branches. Indeed, such a connection exists. At a first glance one would expect to be able to establish a 
 connection between $\alpha^{I}$ and $\alpha^{II}$ or $\beta^{I}$ and $\beta^{II}$. However, a detailed study which 
we reserve for the appendix \ref{app:exemplary shadow transforms} shows that the $\beta$ operators of 
one branch are related to the $\alpha$ operators of the other branch by a shadow transformation. This relation is schematically depicted in figure \ref{fig:shadowbetweenbranches}. 
The shadow transformation is an important concept which we introduce in more detail in section \ref{subsec:real weight c}.
 
	 \subsection{The case  $\frac{m^2}{H^2}=h^2$}
This is a special case for which $\nu=0$ suggesting that there should be only one operator 
  with scaling dimension $\Delta=h$. The analytic expression for the small argument limit of $H^{(1)}_0$ is
	 \be 
\label{hankel10inlimit} 
H^{(1)}_0(z)\approx 1 + \frac{2i}{\pi}\ln(z) + O(z)^2.
\ee
so that \eqref{phibulksoln} leads to
\begin{equation}
  \lim\limits_{q|\eta|\to0}\phi(\vec{x},\eta)
  \rightarrow  
\int\frac{d^{2h}q}{(2\pi)^{2h}} |\eta|^{h}  \left[
 a_{\vec{q}} + a^\dagger_{-\vec{q}} 
+ \frac{2i}{\pi} \ln(q|\eta|) \left(a_{\vec{q}} - a^\dagger_{-\vec{q}} \right) \right]e^{i\vec{q}\cdot\vec{x}}
 \end{equation}
We observe that the first piece is finite and we deduce that the boundary operator is
 \be
\label{nu0boundaryop} 
\alpha^{\nu=0}(q) =  a_{\vec{q}} + a^\dagger_{-\vec{q}} ~~\text{with}~~\Delta^{\nu=0}=h
\ee
The 2nd piece is logarithmically divergent as $\eta\rightarrow 0$. We speculate that one possibility is to impose the additional condition 
$a_q^\dagger = a_{-q}$ which removes the divergence. Otherwise, if the log is kept, it likely leads to 
a second solution which does not fall into the template  \eqref{ltformat}. 
This merging of solutions have been noted to be associated with approach to critical behavior. In the context of AdS/CFT, it has been noticed that such solutions from the gravity side correspond to logarithmic CFT's, see \cite{Gurarie:1993xq} for an introduction of logCFT's based on linear second order differential equations at a critical point and \cite{Grumiller:2013at} for further generalizations.\footnote{We thank Joris Raeymaekers for pointing out the connection with logCFT's.}
	 
\subsection{Heavy Scalars}
	 For scalar fields with mass in the range $\frac{m^2}{H^2}>h^2$, the parameter $\nu$ is purely imaginery, $\nu=i\rho$. With such a parameter,
the Bessel equation\eqref{Bessel_eq}
	 becomes
	 \be 
  \label{imaginerybessel} 
  \frac{d^2}{du^2}\varphi(u)+\frac{1}{u}\frac{d}{du}\varphi(u)+\left(1+\frac{\rho^2}{u^2}\right)\varphi(u)=0
\ee
	 where we previously defined $\phi_q(\eta)\equiv \eta^{h}\varphi(q\eta)$ and $u\equiv q\eta$. The solutions to \eqref{imaginerybessel} are the functions $\tilde{J}_\rho(u)$ and $\tilde{Y}_\rho(u)$, who are related to the solutions $J_\nu(u)$ and $Y_\nu(u)$ of \eqref{Bessel_eq} as follows
	 \begin{subequations}
	 	\begin{align}
	 	\tilde{J}_\rho(u)=\sech\left(\frac{\pi\rho}{2}\right)\Re\left[J_{i\rho}(u)\right]\\
	 	\tilde{Y}_\rho(u)=\sech\left(\frac{\pi\rho}{2}\right)\Re\left[Y_{i\rho}(u)\right]
	 	\end{align}
	 \end{subequations}
	 where $\rho\in \RR$, and $u\in(0,\infty)$. The limiting form of these functions are as follows
	 \begin{subequations}
	 \begin{align}
	 \lim\limits_{u\to 0}\tilde{J}_{\rho}\simeq&\sqrt{\frac{2}{\pi\rho}} \cos\left(\rho \ln\left(\frac{u}{2}\right)-\gamma_\rho\right)\\
	 \lim\limits_{u\to 0}\tilde{Y}_\rho(u)\simeq&\sqrt{\frac{2}{\pi\rho}}\sin\left(\rho \ln\left(\frac{u}{2}\right)-\gamma_\rho\right)
	 \end{align}
	 \end{subequations}
	 where the coefficient $\gamma_\rho$ is defined by
	 	 \be \Gamma(1+i\rho)=\left(\frac{\pi\rho}{\sinh(\pi\rho)}\right)^{1/2}e^{i\gamma_\rho}.\ee
	 
	 In this case the solution that satisfies the Bunch-Davies initial condition is
\be 
\label{complexorderphi}
\phi(\vec{x},\eta) = \int \frac{d^{2h}q}{(2\pi)^{2h}}\eta^{h}\left[
  \tilde{H}^{(1)}_\rho (q|\eta|) a_{\vec{q}}\, e^{i\vec{q}\cdot\vec{x}}
 + \tilde{H}^{(1)*}_\rho (q|\eta|) a^\dagger_{\vec{q}} \, e^{-i\vec{q}\cdot\vec{x}}
\right],
\ee
	 where
	 \be \tilde{H}^{(1)}_\rho (q|\eta|)=\tilde{J}_\rho(q|\eta|)+i\tilde{Y}_\rho(q|\eta|).\ee
Our goal is now to put \eqref{complexorderphi} into  the template given by \eqref{LTL_form} by taking the late time limit.
For this purpose we set again $\vec{q}\to -\vec{q}$ in \eqref{complexorderphi}. The late time limit of the Hankel function works as follows
\begin{subequations}
  \label{latetimehankel}
\begin{align}
\lim\limits_{q|\eta|\to0} \tilde{H}^{(1)}_\rho =& \lim\limits_{q|\eta|\to0} \tilde{J}_\rho(q|\eta|)+i\lim\limits_{q|\eta|\to0}\tilde{Y}_\rho(q|\eta|)
\\
	 	=& \sqrt{\frac{2}{\pi\rho}} \bigg\{ \, \sqrt{\tanh\left(\frac{\pi\rho}{2}\right)} \cos\left[\rho~ \ln\left(\frac{q|\eta|}{2}\right)-\gamma_\rho\right]
\\
	 	\nn 
 &+ i\sqrt{\coth\left(\frac{\pi\rho}{2}\right)}\sin\left[\rho~ \ln\left(\frac{q|\eta|}{2}\right)-\gamma_\rho\right]
\,
\bigg\}.
\end{align}
\end{subequations} 
	 
	 By noting that $\rho=-i\nu$ while $\cos(i\theta)=\cosh(\theta)$ and $\sin(i\theta)=i\sinh(\theta)$, the trigonometric functions can be turned into hyperbolic functions which act to take the inverse of the natural logarithm. Thus one can rewrite the trigonometric functions as
	 \begin{subequations}
  \label{handlingtrig}
	 	\begin{align}
	 \nn \cos\left[\rho~ \ln\left(\frac{q|\eta|}{2}\right)-\gamma_\rho\right]&=\frac{\cos\gamma_\rho}{2}\left[\left(\frac{q|\eta|}{2}\right)^\nu+\left(\frac{q|\eta|}{2}\right)^{-\nu}\right]\\
	 	& -i\frac{\sin\gamma_\rho}{2}\left[\left(\frac{q|\eta|}{2}\right)^\nu-\left(\frac{q|\eta|}{2}\right)^{-\nu}\right]
\\
	 	\sin\left[\rho~ \ln\left(\frac{q|\eta|}{2}\right)-\gamma_\rho\right] &=
   -i\frac{\cos\gamma_\rho}{2}\left[\left(\frac{q|\eta|}{2}\right)^\nu-\left(\frac{q|\eta|}{2}\right)^{-\nu}\right]
\nn
\\
	 	& -\frac{\sin\gamma_\rho}{2}\left[\left(\frac{q|\eta|}{2}\right)^\nu+\left(\frac{q|\eta|}{2}\right)^{-\nu}\right].
	 	\end{align}
	 \end{subequations}
	 Plugging equations \eqref{handlingtrig} into \eqref{latetimehankel} one obtains
	 \be 
  \label{latetimehankel2}
	 \lim\limits_{q|\eta|\to0} \tilde{H}^{(1)}_\rho(q|\eta|) 
  = c_\rho\left[\left(\frac{q|\eta|}{2}\right)^\nu+\left(\frac{q|\eta|}{2}\right)^{-\nu}\right]
+ d_\rho \left[\left(\frac{q|\eta|}{2}\right)^\nu-\left(\frac{q|\eta|}{2}\right)^{-\nu}\right],
	 \ee
	 where the complex coefficients $c_\rho$ and $d_\rho$ are defined as 
	 \begin{subequations}
	 	\begin{align}
	 	c_\rho\equiv &\sqrt{\frac{2}{\pi\rho}} \left[
  \sqrt{\tanh\left(\frac{\pi\rho}{2}\right)}\frac{\cos\gamma_\rho}{2}-i\sqrt{\coth\left(\frac{\pi\rho}{2}\right)}\frac{\sin\gamma_\rho}{2}
\right],
\\
	 	d_\rho\equiv& \sqrt{\frac{2}{\pi\rho}} \left[ \sqrt{\coth\left(\frac{\pi\rho}{2}\right)}\frac{\cos\gamma_\rho}{2}
  -i\sqrt{\tanh\left(\frac{\pi\rho}{2}\right)}\frac{\sin\gamma_\rho}{2}\right].
	 	\end{align}
	 \end{subequations}
	With everything put together and noting that $\nu^*=-\nu$, the late time limit for $\phi(\vec{x},\eta)$ gives
	 \begin{subequations}
  \label{ltmassivephi}
	 \begin{align} \lim\limits_{q|\eta|\to 0}\phi(\vec{x},\eta)=\int\frac{d^{2h}q}{(2\pi)^{2h}}\Big{\{}&\eta^{h+\nu}\left[\left((c_\rho+d_\rho)a_{\vec{q}}+(c^*_\rho-d^*_\rho)a^\dagger_{-\vec{q}}\right)\left(\frac{q}{2}\right)^\nu\right]\\
	 &+\eta^{h-\nu}\left[\left((c_\rho-d_\rho)a_{\vec{q}}+(c^*_\rho+d^*_\rho)a^\dagger_{-\vec{q}}\right)\left(\frac{q}{2}\right)^{-\nu}\right]\Big{\}}.
	 \end{align}
	 \end{subequations}
	 Matching \eqref{ltmassivephi} to \eqref{ltformat} we can read off that for $\nu = i\rho$ and  $\rho>0$
	 \begin{subequations}
	 	\begin{align}
	 \label{massivealpha}	
	 	\alpha(\vec{q})=\left[ \left(c_\rho-d_\rho\right)a_{\vec{q}} + \left(c^*_\rho+d^*_\rho\right)a^\dagger_{-\vec{q}}\right]
  \left(\frac{q}{2}\right)^{-\nu}~~\text{where}~~\Delta_-=h-\nu,\\
	 	\label{massivebeta}
	\beta(\vec{q})=\left[ \left(c_\rho+d_\rho\right) a_{\vec{q}}+ \left(c^*_\rho-d^*_\rho\right)a^\dagger_{-\vec{q}}\right]\left(\frac{q}{2}\right)^\nu~~\text{where}~~\Delta_+=h+\nu.
	 	\end{align}
	 \end{subequations}

	 \section{Elements and Elementary Representations of $G=SO(2h+1,1)$}
	 \label{sec:elementsofG}
	 The main purpose of this section is to introduce how the elementary representations of $SO(2h+1,1)$ can be realized, and set the notation. With this in mind
we summarise some general results from \cite{Dobrev:1977qv} that serve  as a starting point for the following sections. Further details 
 are left for the appendix \eqref{app:furtherpropertiesofG}.
	 
	 \subsection{Group Elements of $SO(2h+1,1)$}
	 The group $G=SO(2h+1,1)$, with $h$ a half-integer, is composed of all linear transformations on the real $2h+2$-dimensional vector space that leave the quadratic form 
	 \be\xi^2=\xi\eta\xi={\xi_1}^2+...+{\xi_{2h+1}}^2-{\xi_0}^2~~\text{where}~~\eta_{\mu\nu}=diag(-1,1,...,1),\ee
	 invariant. The elements $g$ of the group $G=SO(2h+1,1)$ satisfy
	 \be g^T\eta g=\eta,~~\det g=1,~~{g^0}_0\geq1,\ee
	 and the generators $X$ of the corresponding Lie algebra obey
	 \be X^T\eta+\eta X=0.\ee
As a specific example we list the Killing vectors for the case $h=3/2$ (so that $\Ncal=4$) that is of specific interest for cosmological applications, in appendix \ref{sec:killing vectors}. 

	 The group $SO(2h+1,1)$ is composed of six subgroups: the maximal compact subgroup $K = SO(2h+1)$,
 the Euclidean Lorentz group $M=SO(2h)$, the group $N$ of special conformal transformations, the group $\tilde{N}$ of translations, the group  $A=SO(1,1)$ of dilatations
and the Cartan subgroup $H$.
We list these subgroups, typical element notation, generators and irreducible representations of some of them, as relevant in to this work, in table \ref{tablesubgroups}. 
	  Any element $g \in G$ can be decomposed in terms of elements of these subgroups as noted in the appendix \eqref{sec:decompositionofg}.
The subgroup $P = NAM$ is called the parabolic subgroup and plays a central role in the induced representations that we present in the next subsection.
	 \begin{center}
	 \begin{table}
	 \begin{tabular}{|c|p{3.5cm}|c|c|p{2.7cm}|}
	 \hline
	 & $\Ncal = 2h+1$, \; $2h \in \ZZ$ & elements& generators& irreducible representation\\
	 \hline
	 Group & $G=SO(2h+1,1)$ & g & $X_{\mu\nu}$ $\qquad \mu,\nu=0,1,...,2h+1$ & $\chi=[l,c]$\\
	 	\hline
	 	\hline
	 & 	$K=SO(2h+1)$  & k & &  \\
	 & 	 maximal compact subgroup &  &  \; \;~~$ X_{AB} \qquad A,B=1,...,2h+1$ &  \\
	 	\cline{2-5}
	 &	$A=SO(1,1)$ &  &  & \\
	 & noncompact dilatations& a& $D=X_{2h+1~0}$ &$|a|^{-h-c}=|a|^{-\Delta}$\\
	 	\cline{2-5}
	 Subgroups&	$M=SO(2h)$ &  &  & \\
	 & Euclidean Lorentz Group & m & $X_{ij}$     $\quad i,j=1,...,2h$ & $D^l(m)$\\ 
	 	\cline{2-5}
	 &	$ N$ Special Conformal Transformations & n & $C_i=X_{i 0}-X_{i 2h+1}$ &\\
	 \cline{2-5}
	 &$\tilde{N}$ Translations & $\tilde{n}$ & $T_i=X_{i 0}+ X_{i 2h+1}$ & \\
	 \cline{2-5}
	 &	H Cartan subgroup & & &\\
	 	\cline{2-5}
	 \hline
	 \end{tabular}
	 \caption{The subgroups of $SO(2h+1,1)$}
	 \label{tablesubgroups}
	 \end{table}
	 \end{center}
	 
	 The quadratic Casimir of $SO(2h+1,1)$ is
	 \begin{align}\label{eq:casimir} \Omega&=-\frac{1}{2}X^2_{ij}+D^2+2hD+C_i T_i\\
	 &=\ell(\ell+2h-2)+c^2-h^2,\end{align}
	 which depends on the spin $\ell$ and scaling weights $h$ and $c$. Accordingly the irreducible representations are 
labeled by, $\ell$ the label for irreducible representations $D^\ell(m)$ of the subgroup $M=SO(2h)$, $h$ which depends on 
the dimensionality of $G$, i.e. $\dim(G)=2h+1$, and weight $c$ which can be either real or purely imaginary. 
	 
	 \subsection{Induced (Elementary) Representations}

In this subsection we present the induced elementary representations of  $SO(2h+1,1)$ by the parabolic subgroup $P = NAM$, following \cite{Dobrev:1977qv}.
In general, a representation of a group $G$ into a vector space $V$ is a map from $G$ to the group of automorphisms $\aut(V)$ of $V$, that is $\Pi: G \rightarrow \aut(V)$.
An element $g \in G$ is mapped to an element $\Pi_g \in \aut(V)$ with the requirement that the map is a homomorphism, that is for an~\footnote{
	 As an example, consider the $d$-dimensional Hilbert space $\Hcal = {\CC}^d$ of complex vectors endowed with the usual inner product. Consider also
  the set of homogeneous polynomials of degree $\ell$, denoted by 
 $\Pcal_\ell: \Hcal \rightarrow \CC$.  For $p \in \Pcal_\ell$, and $z\in \Hcal$ we have $p(z) = p_{I_{1}\ldots I_{\ell}} z_{I_1} \ldots z_{I_\ell}$ where 
the indices $I_m$ with $m=1 \ldots \ell$ take values in the integer set $1\ldots d$ (summation is understood with repeated indices) and
 $p_{I_{1}\ldots I_{\ell}}\in \CC$. These polynomials act as operators on
$\Hcal$ and  it may be easily shown that $\Pcal_\ell$ is a vector space over $\CC$. If we 
now consider the group $G = GL(d,\CC)$ acting on $\CC^d$, and we define the map $g \rightarrow \Pi_g$ for any $g\in G$ through the left action
\begin{equation}
  (\Pi_g p)(z)  = p ( g^{-1} z) \qquad \forall \quad p\in \Pcal_\ell, \; z \in \Hcal,\; g\in G
\end{equation}
then $ \Pi_{g}[ (\Pi_{g'} p) ] (z)  =  (\Pi_{g'} p)({g}^{-1}z) = p( {g'}^{-1} {g}^{-1} z) = p [( gg' )^{-1} z] = (\Pi_{gg'} p)(z) $ so that \eqref{homomorphism} is satisfied 
and $\Pi_g$ is a representation of $G$ in $\Pcal_\ell$.
%
}
 element $v\in V$
\begin{equation}
  \Pi_g\Pi_{g'} v=\Pi_{gg'} v \qquad \forall \quad g,g' \in G.
\label{homomorphism}
\end{equation}

In constructing the representations of $SO(2h+1,1)$ induced by $P = NAM$, we start from  
 the unitary induced representations of $M$ and $A$ denoted by $D^\ell(m)$ and $|a|^{-(h+c)}$ respectively, where $m\in M$ and $a\in A$ are generic elements of $M$ and $A$.
Taking these together, the unitary induced representations of $P$ are simply $|a|^{-h-c} D^\ell(m)$.
Then the representations of $SO(2h+1,1)$ induced by $P$ are labeled by the pair 
\begin{equation}
\chi = \{\ell, c\}
\end{equation}
 where the spin $\ell$ labels the $M$ representations and the scaling weight $c$  labels the $A$ representations and are constructed as follows.
	 
We start from the Hilbert space $\Vcal^\ell$ which realizes the unitary representations of $M$ and consider 
$\Vcal^\ell$-valued infinitely differentiable functions $f$ on $G$, that is $\mff: G \to \Vcal^\ell$, $\mff(g) \in \Vcal^\ell$.
The functions $\mff$ live on the space $\Ccal_\chi$, i.e. $\mff \in \Ccal_\chi$ and are required to satisfy the so called covariance condition
\begin{equation}
  \mff(gnam) = |a|^{h+c} D^\ell(m)^{-1} \mff(g),
  \label{covariancecond} 
\end{equation}
We now construct the  representations of $G$ induced by  $P$ by taking the space $\aut( \Ccal_\chi)$ of automorphisms of $\Ccal_\chi$  and considering the maps
$\Ical^\chi : G \to \aut( \Ccal_\chi)$, such that $\Ical^\chi_g \in \aut( \Ccal_\chi)$ defined by
\be 
  (\Ical^\chi_g\, \mff)(g')=\mff(g^{-1}g') \qquad \forall \; g,g'\in G,~~\mff\in\Ccal_\chi.
\label{I_chi_reps}
\ee
It is straightforward to show that $\Ical^\chi_g$ obey \eqref{homomorphism}.


For comparison, by the Iwasawa decomposition mentioned in appendix \ref{app:furtherpropertiesofG}, $g=kna$ and in the absence of $m$, the covariance condition of $\Ccal_\chi$  reads
\be 
  \mff(kna)=|a|^{h+c}\mff(k).
\ee
	 This condition defines a space of covariant functions on $K$,
	 \be 
C(K,\Vcal^\ell)=\{\mff(k):K\to \Vcal^\ell; \qquad \mff(kna)=|a|^{h+c}\mff(k)\},
\ee
	 where each space $\Ccal_\chi$ can be identified with a space $C(K,\Vcal^\ell)$. 
 A given irreducible representation $\chi$ of $G$ is related to a unique irreducible representation of K.~\footnote{We 
refer the readers interested more on this matter to sections 2.A and 3.A of \cite{Dobrev:1977qv}.}
	 Finally the compact picture realization of the elementary representation, denoted as $\tilde{\Ical}^\chi$ is 
	 \be 
\left(\tilde{\Ical}^\chi_g\mff\right)(k)=|a|^{h+c}\mff(k_g).
\ee
	 From the $M-$invariant scalar product $\langle,\rangle$ on $\Vcal^\ell$, one can define a $K-$invariant scalar product $(,)$ on $C(K,\Vcal^\ell)$ via
	 \be (\mff_1,\mff_2)=\int_K dk\langle\mff_1(k),\mff_2(k)\rangle\ee  
	 where $dk$ is the normalized Haar measure on $K$. 
	 
	 The possibility to restrict the definition of representations on function space $\Ccal_\chi$ to the compact subspace K suggests that the integral over the compact space will stay finite, leading to a finite inner product which in turn can be recognized as a finite probability rate.
	 
\subsubsection{Connection to functions over the Euclidean space $\RR^{2h}$}
\label{sec:connection to functions over coordinate space}
	 How do the representations realized by functions $\mff \in \Ccal_\chi$ acting on $G$, relate to functions $f :\RR^{2h} \to V^\ell$ 
 that act on the $\vec{x}$-space $\RR^{2h}$?
	 There is a unique correspondence between the elements of Euclidean space $\vec{x}\in \RR^{2h}$, and the elements $\tilde{n}\in\tilde{N}$
  of the subgroup of translations such that the functions over $\RR^{2h}$  match the functions over $\tilde{N}$ via
\be
 \label{fxfn} 
f(\vec{x})=\mff(\tilde{n}_{\vec{x}}).
\ee 
	 Here $\tilde{n}_{\vec{x}}$ denotes the specific element of $\tilde{N}$ that corresponds to the specific element  $\vec{x} \in \RR^{2h}$.
	   
Now $\tilde{n}\in \tilde{N}$ is also related to any $g \in G$, by the Bruhat decomposition $g=\tilde{n}nam$. Considering the representations $\Ical^\chi$ defined by
\eqref{I_chi_reps}  and setting $g'=\tilde{n}_x$
 leads to
\be 
  (\Ical^\chi_g\, \mff)(\tilde{n}_{\vec{x}})=\mff(g^{-1}\tilde{n}_{\vec{x}}) 
\label{eq:rep} 
\ee
	 The element $\vec{x}_g \in \RR^{2h}$ corresponding to a group element $g \in G$ is defined as 
	 \be 
  \label{eq:argument} 
  g^{-1}\tilde{n}_{\vec{x}}=\tilde{n}_{\vec{x}_g}n^{-1}a^{-1}m^{-1}.
\ee
so that \eqref{eq:rep} becomes
\be  
 (\Ical^\chi_g\, \mff)(\tilde{n}_{\vec{x}})= \mff\left(\tilde{n}_{x_g}n^{-1}a^{-1}m^{-1}\right).
\ee 
	 Plugging $g=\tilde{n}_{\vec{x}_g}n^{-1}a^{-1}m^{-1}$ into the covariance condition \eqref{covariancecond} leads to
\be 
 \mff(\tilde{n}_{\vec{x}_g}n^{-1}a^{-1}m^{-1})= |a|^{-h-c} D^\ell(m)\mff( \tilde{n}_{\vec{x}_g}) 
\ee
Now, the set of functions $f$ form the space $C_\chi$, that is $f \in C_\chi$ and with the identification \eqref{fxfn}, one arrives at
\be 
\label{repinxspace} 
  (T^\chi_g\, f)(\vec{x})= |a|^{-h-c} D^\ell(m) f(\vec{x}_g)
\ee
which defines the representations $T^\chi: G \rightarrow \RR^{2h}$ with $T^\chi_g \in \aut(C_\chi)$ of $G$ in $\RR^{2h}$.
	 
To summarize, we discussed how the elementary representations $\chi=\{\ell,c\}$ are realized by infinitely differentiable covariant functions  $\mff \in \Ccal_\chi$ 
and further discussed their properties.  Considering the realization of the representations as functions on the Euclidean space $\RR^{2h}$,
  they are denoted by $f(\vec{x}_g)$ and form the function space $C_\chi$.
The connection between $f(\vec{x}_g)$ and $\mff(g)$ is established through their values on the elements of the subgroup of translations $\tilde{N}$, 
 via  \eqref{fxfn}. We recognize the functions $f(\vec{x}_g)$ as the operators $\alpha$ and $\beta$ identified throughout section \ref{sec:ltandboundaryops} in momentum space.      
that capture the late time behavior of scalar fields of diverse masses  on de Sitter space.

	 \section{Unitary Representations of $SO(2h+1,1)$}
	 \label{sec:unitaryreps}


	 \subsection{Unitarity and the principal series representations}
	 \label{sec:principal series}
	 
	%
	 Now we turn our attention to the $\RR^{2h}$ realization of the representations, $T_g^\chi$. 
 The unitarity of $T_g^\chi$ implies the existence of a bilinear form on $C_\chi$ whose structure is preserved by the representation $T^\chi\otimes T^\chi$ of G. That is
$\forall~f_1,f_2\in C_\chi$ if
\begin{align}
\left(f_1,f_2\right)=&\int \langle f_1(\vec{x}),f_2(\vec{x})\rangle d^{2h}x
\end{align}
then
\begin{align}
\label{unitaritydef}
\left(T^\chi_gf_1,T^\chi_g f_2\right)=&\int\langle f_1(\vec{x}_g),f_2(\vec{x}_g)\rangle d^{2h}x_g.	
	 \end{align}
	 In return unitary representations lead to positive definite probabilities because the scalar product between two states is understood as a probability rate
and the scalar product on the right hand side of \eqref{unitaritydef} is positive definite by definition.   
	 Following \cite{Dobrev:1977qv} let us work out the condition that this definition brings about for the representations.
	 
	 The elementary representation $\chi=\{\ell,c\}$ is realized on $\RR^{2h}$ via \eqref{repinxspace}. The bilinear product works as
	 \be \left(T^\chi_gf_1,T^\chi_gf_2\right)=\int\langle T^\chi_gf_1,T^\chi_gf_2\rangle d^{2h}x.\ee
	 The right hand side can be expanded by \eqref{repinxspace} as
	 \be \rhs=\int \langle |a|^{-h-c}D^\ell(m)f_1(\vec{x}_g),|a|^{-h-c}D^\ell(m)f_2(\vec{x}_g)\rangle d^{2h}x.\ee
	 Since $\langle Au,Bv\rangle=\langle u,A^\dagger Bv\rangle$ for arbitrary linear operators $A,B$,
	 \begin{align} 
\rhs &=\int  \langle f_1(\vec{x}_g),\left[|a|^{-h-c}D^\ell(m)\right]^\dagger |a|^{-h-c}D^\ell(m)f_2(\vec{x}_g)\rangle d^{2h}x\\
	 	&=\int  |a|^{- 2h-c^*-c} \langle f_1(x_g),{D^\ell(m)}^\dagger D^{\ell}(m)f_2(x_g)\rangle d^{2h}x
\end{align}
where we used the fact that $h \in \RR$ and that $a$ being an element of dilatations is simply a scale factor and can be taken out of the inner product.
	 As $D^\ell(m)$ is an element of $SO(2h)$ then ${D^\ell(m)}^\dagger= {D^\ell(m)}^{T}$ and more over ${D^\ell(m)}^{T}D^\ell(m)=\mathbb{I}$ is the identity element. Therefore
	 \be \rhs=\int \langle f_1(x_g),f_2(x_g)\rangle |a|^{-(c^*+c)}|a|^{-2h}d^{2h}x.\ee
	 The connection between $\vec{x}$ and $\vec{x}_g$ involves the action of special conformal transformations, dilatations and rotations via \eqref{eq:argument}.
	 Among all these transformations only the Jacobian of dilatations is nontrivial and gives
	 \be |a|^{-2h} d^{2h}x= d^{2h}x_g,\ee
so that we arrive at
	 \be \label{innerproductint} \int \langle f_1(x_g),f_2(x_g)\rangle |a|^{-(c^*+c)}dx_g.\ee
	 This matches \eqref{unitaritydef} provided $c^*=-c$, that is if $c$ is purely imaginary $c=i\rho$.
	 
	 The function space $C_\chi$ can be completed into a Hilbert space $\Hcal_\chi$ by equipping it with the scalar product \eqref{unitaritydef}. From now on the 
representation $T^\chi$ for $\chi=\{\ell ,i\rho\}$ is identified with its extension to a unitary representation of G in $\Hcal_\chi$. 
This family of unitary representations, constructed via the scalar product \eqref{unitaritydef}, are called the ``\emph{(unitary) principal series} representations'' and 
they are irreducible \cite{hirai1962_1,hirai1962_2}.
	 
	 The heavy scalars, that is, with masses $\frac{m^2}{H^2}>h^2$ accommodate boundary operators with purely imaginary weight
 among the late time solutions of section \ref{sec:ltandboundaryops}. Therefore the operators \eqref{massivealpha} and \eqref{massivebeta} belong to the principal series representations.
	 
\subsection{Real weight $c$ and the complementary series}
	 	\label{subsec:real weight c}
We now turn to the case where the weight $c$ is a real number.  Remember that the subtlety was that the volume element in \eqref{innerproductint} 
 contains the factor $|a|^{-(c*+c)}$ which is non-vanishing for real $c$.
 Yet if there exists an operator $A$ such that $\tilde{f}(x)=[Af](x)$ has $\tilde{c}=-c$ for each $f(\vec{x})$ of weight $c$, 
 then the inner product $\left(f,Af\right)$ will involve $|a|^{-(c+\tilde{c})}=|a|^{-(c-c)}$ and lead to a unitary representation. 
Such an operator $A$ indeed exists. It is defined via similarity transformations \cite{Wigner1902} and 
 is referred to as an \emph{intertwining operator} in early works \cite{Dobrev:1977qv}, or $[Af](x)$ is referred 
to as the \emph{shadow transformation} in more recent works \cite{Anninos:2017eib}. Below we explore how the shadow transformation is expressed and how it works.  

\subsubsection{Definition of the Intertwining operator}
We are interested in the  \emph{character} of a representation $\Pi_g$ defined simply by its trace $\Tr \Pi_g$. 
A theorem stated in the first paragraph of section 4.A. of  \cite{Dobrev:1977qv} (with reference to \cite{Warner_book})
 states that every $K$-finite unitary representation of $G$ is determined uniquely by the character of the representation, up to equivalence.
Any irreducible representation $\ell$ of $SO(2h+1)$ is equivalent to its mirror image 
$\tilde{\ell}$, where for $\chi=\{\ell,c\}$ its mirror image is an $O(\ndim)$  transformation that 
includes reflections $S$ such that
	 	\be 
  S^2=S^{T}S=1,~~\det S=1
\ee
and act on $D^\ell$ the irreducible representations of $SO(\ndim)$ as
	 	\be D^{S \ell}(\Lambda)\equiv D^\ell (S\Lambda S^{-1}),\ee
	 	leading to  $\tilde{\chi}=\{\tilde{\ell},-c\}$\footnote{The proof of this relation is given in Corollary 3.3 of \cite{Dobrev:1977qv}.}. 
For $\underline{\ell}= (\ell_1,....,\ell_{[h]})$
  the mirror image corresponds to $\underline{\tilde{\ell}}=\left( (-1)^{2(h-[h])}\ell_1,\ell_2,...,\ell_{[h]}\right)$,
where $[h]$ denotes the integer value of $h$ rounded down, e.g. $[3/2] = 1$.
 For a scalar representation $\underline{\ell}=(0,...,0)$ and hence $\tilde{\ell}=\ell$.~\footnote{
The notation $\underline{\ell}= (\ell_1,....,\ell_{[h]})$ denotes the highest weight representation of the group $M=SO(2h)$ where the integer part of $h$, $[h]$ denotes
the rank of the group. The labels $\ell_i$ have a hierchical order. This order is determined by the odd or even dimensionality of the group. Emphasizing the rank of the group
\begin{itemize}
 \item for $SO(2[h])$ the ordering is $|\ell_1|\le \ell_2 \le \ldots \le \ell_{[h]}$
 \item for $SO(2[h]+1)$ the ordering is $0\le\ell_1\le \ell_2 \le \ldots \le \ell_{[h]}$
\end{itemize}
As an example in the case of  $dS_4$ we have $h=3/2$ and $M = SO(3)$ which has rank $[h]=1$. Thus since $3$ is odd the highest weight representation is $\underline{\ell}=(\ell_1)$
where $\ell_1\ge 0$.  In the case of $dS_5$ we have $h=2$ and $M=SO(4)$ which has rank $[h]=2$. Then, since $4$ is even the highest weight representation in this case is 
$\underline{\ell}= (\ell_1, \ell_2)$ where $|\ell_1|\le \ell_2$.
}
	 	
	%
	 	 Thus, at this point we have gained awareness of the equivalence of representations $\chi=\{\ell,c\}$ and $\tilde{\chi}=\{\tilde{\ell},-c\}$. 
For two equivalent representations $T$ and $T'$ acting on $\Hcal_\chi$ and $\Hcal'_{\chi}$ respectively, there exist a continuous linear 
map\footnote{In general, representations related to each other by a similarity transformation, which is what
 equation \eqref{eqandA} is, are called equivalent and because the trace is not effected by the similarity transformation, equivalent representations have the same trace.}
	 	\be \label{eqandA} A:\Ccal_\chi\to \Ccal'_\chi~~\text{such that}~~AT_g=T'_gA~~\text{for all}~~g\in G.\ee
	 	This linear map A is called the intertwining operator. 
	 	Strictly speaking, the relation \eqref{eqandA} on its own guarantees partial equivalence. For equivalent representations, such as the $\chi$ and $\tilde{\chi}$ of interest, A also has a continuous inverse. Since the intertwining map $A$, relates a representation $\chi$ with weight $c$ to a representation $\tilde{\chi}$ with weight $\tilde{c}=-c$, the action of this map guarantees that $(f, Af)$ gives a unitary inner product. In other words while $(f,f)$ is not unitary, the inner product $(f,\tilde{f})$, where $\tilde{f}=Af$, is.
	 	
Consider now $\tilde{\mff} \in \Ccal_{\tilde{\chi}}$ on $G$ as defined in the previous section. This realizes the IR $\tilde{\chi}=\{\tilde{\ell},\tilde{c}\}=\{\tilde{\ell},-c\}$ and 
hence satisfies the covariance condition
\begin{align} 
	 		\nn 
    \tilde{\mff}(gnam)&=|a|^{h+\tilde{c}}D^{\tilde{\ell}}(m)^{-1} \tilde{\mff}(g)
\\
	 		\label{covtilde} &=|a|^{h-c}D^{\tilde{\ell}}(m)^{-1} \tilde{\mff}(g).
\end{align}
The intertwining operator $A_\chi$ is a map 
	 	\be
  \label{mapAX} A_\chi:\Ccal_{\tilde{\chi}}\to\Ccal_{\chi}~~\text{which is well defined and analytic for}~~ \Re(c)<0,\ee
	 	and is expressed as~\cite{Dobrev:1977qv}
	 	\begin{align} 
  \label{int1} \left(A_\chi \tilde{\mff}\right)(g)
 &= \int_{\tilde{N}}\tilde{\mff}(gw\tilde{n}_x)d^{2h}x
\\
 &= 
  \int_{\tilde{N}}\tilde{\mff}(gwk_x)\frac{d^{2h}x}{\left(1+x^2\right)^{h-c}}
	 	\label{int1line2}	
\end{align}
where $x^2 = |\vec{x}|^2$ and where $w$ is an element of the Weyl group  $W = M'/M$ where $M' \cong O(2h)$ is the normalizer of $A$ in $K$, i.e. the set of elements $m' \in K$ such that $m' a {m'}^{-1} \in A,\;\; \forall a \in A$.

The inverse map is given by
	 	\be A_{\tilde{\chi}}:\Ccal_\chi\to\Ccal_{\tilde{\chi}}~~\text{which is well defined and analytic for}~~\Re(c)>0,\ee
	 	and in comparison with equation \eqref{mapAX} it can be defined as
	 	\be [A_{\tilde{\chi}}\mff](g)=\int_{\tilde{N}}\mff(gwk_x)\frac{d^{2h}x}{\left(1+x^2\right)^{h+c}}~~\text{where}~~\mff\in \Ccal_\chi.\ee	 
	 	Among the representations $\ell$ of $M$, those who are equivalent to their mirror representations\footnote{Even though the mirror image representation $\tilde{\ell}$ 
is equivalent to the original representation $\ell$, the matrices $D^{\tilde{\ell}}$ and $D^{\ell}$ are different except for the case $\ell=0$.} $\tilde{\ell}=\ell$,
 make up a special case that can be extended to representations of $SO(2h+1,1)$. For this special case the equivalence map between $\{\ell,-c\}$ and $\{\tilde{\ell},-c\}$ can be defined by
	 	\begin{align}
	 		\Ical(I_s):\Ccal_{\{\tilde{\ell},-c\}}\to{C}_{\{\ell,-c\}}~\text{such that}~[\Ical(I_s)\mff]=D^{\ell}(I_s)\mff(g)
	 	\end{align}
	 	where $I_s(x_{1\ldots 2h-1},x_{2h})=(-x_{1\ldots 2h-1},x_{2h})$ is the reflection.
 Then the normalized intertwining operators are denoted by $G_\chi:\Ccal_{\tilde{\chi}}\to\Ccal_\chi$ and $G_{\tilde{\chi}}:\Ccal_\chi\to \Ccal_{\tilde{\chi}}$ and are given as
	 	\begin{align}
	 	\label{normalizedGX}	G_\chi:\Ccal_{\tilde{\chi}}\to\Ccal_\chi~~&\text{where}~~G_\chi=\gamma_\chi A_\chi\Ical(I_s)\\
	 	G_{\tilde{\chi}}:\Ccal_\chi\to \Ccal_{\tilde{\chi}}~~&\text{where}~~G_{\tilde{\chi}}=\gamma_{\tilde{\chi}}\Ical(I_s)A_{\tilde{\chi}}.
	 	\end{align}
	 	Here $\gamma_\chi$ is a normalization factor that is to be determined so that the normalized intertwining operators obey the normalization condition
	 	\be\label{normalizationofG} G_\chi G_{\tilde{\chi}}=1=G_{\tilde{\chi}}G_\chi.\ee
	 	
	 	There is some subtlety in obtaining $G_\chi$ from $A_\chi$ of equation \eqref{int1line2} via equation \eqref{normalizedGX}, due to the equivalence map $\Ical\left(I_s\right)$. Leaving the details to appendix \ref{app:normGX2}, we quote the result of \cite{Dobrev:1977qv} for the normalized intertwining operator, as it acts on real space functions, for the irreducible representations of $SO(2h+1,1)$. The normalized $G_\chi$ acts on functions $\tilde{f}\in C_{\tilde\chi}$ as
	 	\begin{align}
	 		\left[G_\chi\tilde{f}\right](\vec{x})&=\int G_\chi(\vec{x}-\vec{y})\tilde{f}(\vec{y}\,)d^{2h}y\\
	 		\label{intop}&=\int\frac{\gamma_\chi}{|\vec{x}-\vec{y}|^{2(h+c)}}D^\ell(r({\vec{x}-\vec{y}}))\tilde{f}(\vec{y}\,)d^{2h}y
	 	\end{align}
	 	where $r({\vec{x}-\vec{y}})=m(R,\vec{x}-\vec{y})$ is an $O(2h)$ transformation and $R$ is conformal inversion.
	 	
	 	Via the normalized intertwining operator $G_\chi$, an inner product that respects unitarity can be constructed for the representations with real $c$ as follows
	 	\be(f_1,G_{\tilde{\chi}} f_2)=\int \int \langle f_1(\vec{x}_1),G_{\tilde{\chi}}(\vec{x}_{12})f_2(\vec{x}_2)\rangle d^{2h}x_1 d^{2h}x_2.\ee
where $\vec{x}_{12} = \vec{x}_1 - \vec{x}_2$.
	 	The form $G_{\tilde{\chi}} f_2$ is Hermitian.
	 	\subsubsection{Normalization of the Intertwining operator}
	 	The normalization condition \eqref{normalizationofG} $G_\chi G_{\tilde{\chi}}=1=G_{\tilde{\chi}}G_\chi$, does not 
 uniquely determine the normalization factor $\gamma_\chi$. 
 In past works, four conventions for choosing $\gamma_\chi$ have been used, each of which is useful for a different purpose (ie. the convenient choice of $\gamma_\chi$ 
for Wightman positivity which is appropriate for Minkowski spacetime is different than the choice that is employed for the derivation of operator product expansion in quantum field theory). Here we quote the result of \cite{Dobrev:1977qv} on the appropriate normalization for the positivity of the scalar product  $(f_1,G^+_{\chi}f_2)$ where in momentum space
	 	\be \label{normalizedG} G^+_{\chi}(q)=\left(\frac{q^2}{2}\right)^c\sum^\ell_{s=0}K_{\ell s}(c)\Pi^{\ell s}(q), \ee
	 	and refer the reader to section 5.C of \cite{Dobrev:1977qv} for further details. Here $\ell=0,1,2,..$ denotes 
the spin of the representation under consideration. The coefficient $K_{\ell s}(c)$ is defined as
	 	\be K_{\ell s}(c)=(-1)^{\ell-s}\frac{\Gamma(h+c+s-1)\Gamma(c+2-h-s)}{\Gamma(h+c+\ell-1)\Gamma(c+2-h-\ell)},\ee
	 	and $\Pi^{\ell s}(q)$ are $SO(2h-1)_q$ projection operators that map $\Vcal^\ell_{(2h)}$ onto the subspace $\Vcal^s_{(2h-1)}$,
 where $SO(2h-1)_q$ is the stability group of q with respect to which the harmonic analysis is carried out and the $\Vcal^s_{(2h-1)}$ is the space of $SO(2h-1)$ symmetric, 
traceless tensors of rank $s\leq \ell$. In general these operators can be written in terms of zonal spherical 
functions $C^{h-\frac{3}{2}}_s(z_1,z_2)$ of $SO(2h-1)$\footnote{for the definition of these functions we refer the reader to the appendix A.2 of \cite{Dobrev:1977qv}.}
	 	\be \Pi^{\ell s}(q;z_1,z_2)=A_{\ell s}(-1)^s\left(\frac{(qz_1)(qz_2)}{\frac{q^2}{2}}\right)^\ell C^{h-\frac{3}{2}}_s(\omega) 
\label{P_ell_s_def}\ee
	 	where $A_{\ell s}$ is a normalization constant which guarantees that 
	 	\be \Pi^{\ell s}(q)\Pi^{\ell s'}(q)=\delta_{ss'}\Pi^{\ell s}(q),\ee
	 	and $\omega=\cos\theta =1-\frac{q^2(z_1z_2)}{(qz_1)(qz_2)}$. The case of $2h=3$, which is relevant for $dS_4$ is special. In this case
	 	\be \Pi^{\ell s}_{(2h=3)}(q,z_1,z_2)=(-1)^s\frac{2\ell!}{(\ell-s)!(\ell+s)!}\frac{\Gamma(\ell+\frac{1}{2})}{\Gamma(\frac{1}{2})}\left(\frac{(qz_1)(qz_2)}{\frac{q^2}{2}}\right)^\ell\cos(s\theta).\ee
	 	For a scalar field on $dS_4$, $K_{00}=\Pi^{00}=1$ is the only term that contributes to the sum in \eqref{normalizedG}.
	 	
	 	In summary, quoting the theorem 5.1 of \cite{Dobrev:1977qv}, the inner product
	 	\begin{align} \left(f_1,G^+_\chi f_2\right)&=\int \int \langle f_1(x_1),G^+_\chi (x_{12})f_2(x_2)\rangle d\vec{x}_1 d\vec{x}_2\\
	 		&=\int \langle f_1(q),G^+_\chi(q)f_2(q)\rangle\frac{d^{2h}q}{(2\pi)^{2h}},\end{align}
	 	which is defined on $C_{\tilde{\chi}}\times C_{\tilde{\chi}}$ with the intertwining operator $G^+_\chi$ given by \eqref{normalizedG} 
	 	is positive definite for
	 	\begin{subequations} 
  \label{positivedefreps}
	 		\begin{align}
	 			\ell=0~~~-h&<c<h~~~~~~~(h\geq 1),\\
	 			\ell=1,2,...~~ 1-h&<c<h-1~~(h>1).
	 		\end{align}
	 	\end{subequations}
	 	The unitary representations $\chi$ and $\tilde{\chi}$ of $G=SO(2h+1,1)$ in the domain \eqref{positivedefreps} are equivalent, and they differ only in the sign of $c$. Among the representations of this domain, the ones with $c\neq 0$ are called the \emph{`` complementary series''} of type I unitary representations of $G=SO(2h+1,1)$.
	 	
	 	Among the late time solutions studied in section \ref{sec:ltandboundaryops}, we saw that light scalars, defined to be in the range
\be 
\frac{m^2}{H^2} \; < \;  h^2 ,
 \ee
	 	have real weight $c$. Therefore the boundary operators \eqref{boundaryoperators} that correspond to scalars with mass $\frac{m^2}{H^2}=h^2-\frac{1}{4}$, 
and the boundary operators \eqref{branchrevg0boundary} and \eqref{-2n+1/2branch} that capture light scalars in more generality, are among the complementary series representations 
of $SO(2h+1,1)$. The operators \eqref{boundaryoperators} correspond to representations with $c=\pm \frac{1}{2}$, 
the operators \eqref{branchrevg0boundary} have $c=\nu=|\sqrt{h^2-\frac{m^2}{H^2}}|$, and the operators \eqref{-2n+1/2branch} have $c=\pm \frac{2n+1}{2}$ with $n=0, 1, 2,\dots$. 
	 	
	 \section{The positive definite norm and unitarity}

	 \label{sec:positivenormandunitarity}
	 Now that  we understand the necessity of the intertwining operator, how it works and that we have some expressions for the late time boundary operators at hand, we are ready to check that these operators have positive definite norm and they are unitary representations of $SO(2h+1,1)$.

 \subsection{Example case: Conformally Coupled Scalar field on $dS_4$}
 \label{subsec:conformally coupled scalar}
 Now let us check the positive definiteness of the norm of the operators associated with a conformally coupled scalar field on $dS_4$, whose mass is $m^2=2H^2$. For this case, since $\nu^2=h^2-\frac{m^2}{H^2}=\frac{1}{4}$, $\Delta=h+\nu=\frac{3}{2}\pm \frac{1}{2}$ and in accordance with equation \eqref{ltformat}, the late time solution for this field is of the form
 \be \phi(\eta,\vec{x})=\alpha(\vec{x})\eta^{\Delta_\alpha}+\beta(\vec{x})\eta^{\Delta_\beta}.\ee
 Here the operator $\alpha$ has scaling dimension \be \Delta_\alpha=\frac{3}{2}-\sqrt{\frac{9}{4}-\frac{m^2}{H^2}}=1~~\text{which means}~~c_\alpha=-\frac{1}{2}\ee
  and the operator $\beta$ has
  \be \Delta_\beta=\frac{3}{2}+\sqrt{\frac{9}{4}-\frac{m^2}{H^2}}=2~~\text{which means}~~c_\beta=\frac{1}{2}.\ee
 
 Of course there is also the possibility to use the shadow transforms of these operators with
 \be \tilde{\Delta}_\alpha=3-\Delta_\alpha=2,~~\tilde{c}_\alpha=\frac{1}{2},\ee 
 \be \tilde{\Delta}_\beta=3-\Delta_\beta=1,~~\tilde{c}_\beta=-\frac{1}{2}.\ee
 
As was obtained in section \ref{subsection:special mass soln}
\be \label{alphak}\alpha(\vec{q})= \frac{1}{\sqrt{2q}}\left(a_{\vec{q}}+a^\dagger_{-\vec{q}}\right),\ee
\be \label{betak} \beta(\vec{q})=-i\sqrt{\frac{q}{2}}\left(a_{\vec{q}}-a^\dagger_{-\vec{q}}\right).\ee
We are interested in the normalization of $\alpha(\vec{x})$ and $\beta(\vec{x})$. Since all of these operators have real $c$, in doing this exercise we will see the necessity of the intertwining operator in calculating the norm.

Remember that there exists two types of intertwining operators $G_\chi$ and $G_{\tilde{\chi}}$, where each is well defined over a different range 
that depends on the sign of $c$, and act on a different function space
\begin{align}
G_\chi&:C_{\tilde{\chi}}\to C_\chi,~~\text{for}~~c<0,\\
G_{\tilde{\chi}}&: C_{\chi}\to C_{\tilde{\chi}},~~\text{for}~~c>0.
\end{align}
So the first step for calculating the norm relies on deciding which of the intertwining operators act on the operator of interest. Considering our operators associated with the late time boundary, we have two families, $c_\alpha=-\frac{1}{2}$ and $c_\beta=\frac{1}{2}$. 
  
  The family $c_\alpha=-\frac{1}{2}$ contains
  \be \label{calphafamily}
  \Big{\{}c_\alpha=-\frac{1}{2}\Big{\}}=\{\alpha(\vec{x})\in C_\chi\}.\ee	
  Since for this family $\Re(c_\alpha)<0$, the well defined intertwining operator that first comes to mind is $G^+_{\chi}$,
  \be G^+_{\{0,c\}}(q)=\left(\frac{q^2}{2}\right)^c\ee
   which acts on $\tilde{\alpha}\in C_{\tilde{\chi}}$. But we have the expression for $\alpha\in C_{\chi}$, not for $\tilde{\alpha}$. On the other hand, we know that the intertwining operator
   \be \label{Gforalphaexamp} G_{\tilde{\chi}}:C_{\chi}\to C_{\tilde{\chi}},~\text{where}~G^+_{\{0,\tilde{c}\}}(q)=\left(\frac{q^2}{2}\right)^{\tilde{c}}\ee
   can act on the operator $\alpha(\vec{q})$. It is straight forward to see that, for a scalar field $K_{00}(c)=K_{00}(\tilde{c})=1$ and hence equation \eqref{normalizedG} with $c$ replaced by $\tilde{c}$ gives \eqref{Gforalphaexamp}. For the case at hand
   \be G^+_{\{0,-\tilde{\frac{1}{2}}\}}(q)= \left(\frac{q^2}{2}\right)^{\frac{1}{2}}=G^+_{\{0,\frac{1}{2}\}},\ee
   and its action gives the shadow operator $\tilde{\alpha}(q)\in C_{\tilde{\chi}}$
   \begin{align}
   \nn \tilde{\alpha}(\vec{q})&=G^+_{\{0,\frac{1}{2}\}}\alpha(\vec{q})\\
  \nn &=\frac{q}{\sqrt{2}}\frac{1}{\sqrt{2q}}\left(a_{\vec{q}}+a^\dagger_{-\vec{q}}\right)
\\
   &=\frac{\sqrt{q}}{2}\left(a_{\vec{q}}+a^\dagger_{-\vec{q}}\right).
   \end{align}

   Hence the well defined inner product for the $c_\alpha=-\frac{1}{2}$ family is
    \begin{align} 
  \left(\alpha,G^+_{\{0,-\tilde{\frac{1}{2}}\}}\alpha\right)= \left(\alpha,\tilde{\alpha}\right).
    \end{align}
    Making use of \eqref{alphak} and the Bunch-Davies vacuum state $|0\rangle$ which is annihilated by the operator $a$ such that $a|0\rangle=0$, we can define the ket
    \begin{align}
   \nn |\alpha(\vec{q})\rangle &= \alpha(\vec{q})|0\rangle\\
   \nn &=\frac{1}{\sqrt{2q}}\left(a_{\vec{q}}+a^\dagger_{-\vec{q}}\right)|0\rangle\\
    &=\frac{1}{\sqrt{2q}}|-\vec{q}\rangle.
    \end{align}
    Similarly
    \be |\tilde{\alpha}(\vec{q})\rangle =\frac{\sqrt{q}}{2}|-\vec{q}\rangle.\ee

Defining the volume of momentum eigenstates $\Omega$ by
\begin{equation}
\Omega\equiv \int \frac{d^{2h}q}{(2\pi)^{2h}}  \langle \vec{q} | \vec{q}\rangle 
\end{equation}
we can compute the ``densitized'' inner product
    \begin{align} 
\nn   \frac{1}{\Omega}\left(\alpha,G^+_{\{0,-\tilde{\frac{1}{2}}\}}\alpha\right)  &= \frac{1}{\Omega} \left(\alpha,\tilde{\alpha}\right) \\
   \nn &=  \frac{1}{\Omega}  \int \frac{d^{2h}q}{(2\pi)^{2h}}\frac{1}{\sqrt{2q}}\langle -q|\frac{\sqrt{q}}{2}|-q\rangle\\
    &=\frac{1}{2\sqrt{2}} 
        \end{align}
        which is positive definite. 
 As there is no extra $q$-dependence in $(\alpha,\tilde{\alpha})$, $\alpha$ is normalized up to the normalization of momentum eigenstates.
\footnote{
  Note that had we not used the intertwining operator, the ``densitized'' inner product
  \begin{align}
 \nn \frac{1}{\Omega} \left(\alpha,\alpha\right)&=  \frac{1}{\Omega} \int \frac{d^{2h}q}{(2\pi)^{2h}}\frac{1}{\sqrt{2q}}\langle -q|-q\rangle \frac{1}{\sqrt{2q}}\\
  \end{align}
would be divergent.
}
  
  For the family $c_\beta=\frac{1}{2}$
  \be \Big{\{}c_\beta=\frac{1}{2}\Big{\}}=\{\beta(x)\in C_{\chi}\}\ee
  the weight is positive, i.e. $c_\beta>0$. The appropriate intertwining operator for the shadow transformation is $G^+_{\{0,\tilde{c}\}}(q)$ and the well defined inner product is
  \be \left(\beta,G^+_{\{0,\tilde{\frac{1}{2}}\}}\beta\right)~~\text{with}~~G^+_{\{0,\tilde{\frac{1}{2}}\}}(q)=\frac{\sqrt{2}}{q}.\ee
  The shadow transformed operator here is
  \begin{align}
  \nn \tilde{\beta}&= G^+_{\{0,\tilde{\frac{1}{2}}\}}(q)\beta(\vec{q})\\
 \nn &=-i\frac{\sqrt{2}}{q}\sqrt{\frac{q}{2}}\left(a_{\vec{q}}-a^\dagger_{-\vec{q}}\right)\\
  &=-\frac{i}{\sqrt{q}}\left(a_{\vec{q}}-a^\dagger_{-\vec{q}}\right)
  \end{align}
 We can define a ket associated with $\beta$ as we did earlier on
  \begin{subequations}
  \begin{align}
 \label{statebeta}|\beta(\vec{q})\rangle&=i\sqrt{\frac{q}{2}}|-\vec{q}\rangle\\
 \label{statebetatilde}|\tilde{\beta}(\vec{q})\rangle &=\frac{i}{\sqrt{q}}|-\vec{q}\rangle.
  \end{align}
  \end{subequations}  
 Finally the inner product is
 \begin{align}
 \nn \frac{1}{\Omega}  \left(\beta, G^+_{\{0,\tilde{\frac{1}{2}}\}}\beta\right)&= \frac{1}{\Omega} \int \frac{d^{2h}q}{(2\pi)^{(2h)}} \langle \beta(\vec{q})|\tilde{\beta}(\vec{q})\rangle\\
\nn &= \frac{1}{\Omega}  \int \frac{d^{2h}q}{(2\pi)^{(2h)}} (-i)\sqrt{\frac{q}{2}}\langle -\vec{q}|-\vec{q}\rangle\frac{i}{\sqrt{q}}\\
&= \frac{1}{\sqrt{2}}
 \end{align} 
 which is positive definite.
 
 In summary, for operators with real weight $c$, positive definite norms require the use of the intertwining operator. In return, the presence of the intertwining operator guarantees the unitarity of the representation. The form of the intertwining operator is sensitive to the sign of the real part of the weight $c$. In addition one must pay attention to 
which function space the intertwining operator acts on, as $G^+_{\{\ell,c\}}$ and $G^+_{\{\tilde{\ell},\tilde{c}\}}$ act 
on different function spaces. Lastly, note that everything in this calculation is determined by $c$. Although we considered conformally coupled scalars on $dS_4$ here, our results would be the same for $dS_3$. On $dS_3$ the form for the operators $\alpha^{dS_3}(q)$ and $\beta^{dS_3}(q)$ would not change, it would still be that $c^{dS_3}_{\alpha}=-\frac{1}{2}=c_\alpha$ and $c^{dS_3}_{\beta}=\frac{1}{2}=c_\beta$. The discussion in the choice of intertwining operators, and their form would stay the same as well.\footnote{In the definition of the intertwining operator (see \eqref{normalizedG}) $K_{ls}(c)$ and $\Pi^{ls}(q)$ have implicit $h$-dependence. For scalars, $l=0$ and $K_{00}(c)=1$ for all $h$. The form of $\Pi^{ls}(q)$ is in general dictated by $h$, but $\Pi^{00}(q)=1$ for both cases of $h=\frac{3}{2}$ and $h=1$. This is why the discussion for conformally coupled scalars on $dS_3$ parallels that of $dS_4$.} Only the total dimension would change, since for this case $h=1$, giving $\Delta^{dS_3}_{\alpha}=1+c_{\alpha}=\frac{1}{2}$ and $\Delta^{dS_3}_{\beta}=1+c_{\beta}=\frac{3}{2}$.

\subsection{Example case: Mass $\frac{m^2}{H^2}=h^2$}
This special case with $c=\nu=0$ has the single boundary operator
\be 
\alpha^{\nu=0}(q)\sim a_{\vec{q}}+a^\dagger_{-\vec{q}} ~~\text{with}~~\Delta^{\nu=0}=h.
\ee
As this operator has no $q$-dependence, the intertwining operator is trivial 
\be G^+_{\{0,0\}}(q)=1.\ee
This means the boundary operator $\alpha^{\nu=0}$ is equal to its shadow 
\be \alpha^{\nu=0}(q)=\tilde{\alpha}^{\nu=0}~~\text{and}~~\Delta^{\nu=0}=\tilde{\Delta}^{\nu=0}=h.\ee
It also means that the finite scalar product resembles that of a principal series representation and is simply $(\alpha^{\nu=0},\alpha^{\nu=0})$.
 
 \subsection{Example case: A Heavy Scalar on $dS_4$} 
	 	 
	 	As was discussed in section 4.3, the heavy fields fall in the principal series representations and for the boundary operators of section 4.3, with $c=\nu=i\rho$, there is no need to include any intertwining operators. The coordinate momenta q-dependence of these boundary operators have the form $|\alpha\rangle\sim q^\nu= q^{i\rho}$. Hence $\langle\alpha|\sim q^{\nu^*}=q^{-i\rho}$ which guarantees the q-dependence of the integrand, $\langle \alpha|\alpha\rangle$, cancels itself automatically, leaving only the integration over a Dirac delta function. 
	 	
	 	More explicitly, as was found in equation \eqref{massivebeta} one of the boundary operators is
	 	\be	\beta(\vec{q})=\left[\left(c_\rho+d_\rho\right) a_{\vec{q}}+\left(c^*_\rho-d^*_\rho\right)a^\dagger_{-\vec{q}}\right]\left(\frac{q}{2}\right)^\nu~~\text{where}~~\Delta_+=h+\nu.\ee
	 	 The corresponding ket for this operator is 	 
	 	 \be |\beta(\vec{q})\rangle=\left(\frac{q}{2}\right)^\nu \left(c^*_\rho-d^*_\rho\right)|-\vec{q}\rangle.\ee
	 	 With
	 	 \be \langle\beta(\vec{q})|=\left(\frac{q}{2}\right)^{-\nu} \left(c_\rho-d_\rho\right)\langle-\vec{q}|\ee
	 	 the product $\langle \beta(\vec{q})|\beta(\vec{q})\rangle$ is 
	 	 \begin{align} 
\nn
\langle\beta(\vec{q})|\beta(\vec{q})\rangle&=\left(\frac{q}{2}\right)^{-\nu} \left(c_\rho-d_\rho\right)\langle-\vec{q}|-\vec{q}\rangle
   \left(\frac{q}{2}\right)^\nu \left(c^*_\rho-d^*_\rho\right)\\
	 	 &=|c_\rho-d_\rho|^2 \langle-\vec{q}|-\vec{q}\rangle  
	 	 \end{align}
	 	 Plugging this into the expression for the inner product leads to
	 	 \begin{align} 
  \nn
\frac{1}{\Omega} \left(\beta,\beta\right)&= \frac{1}{\Omega} \int \frac{d^{2h}q}{(2\pi)^{2h}}\langle\beta(\vec{q})|\beta(\vec{q})\rangle\\
	 	 &=|c_\rho-d_\rho|^2
	 	 \end{align}
	 	 which is positive definite.
	 	 
	 	 Lastly the second boundary operator for a heavy field is by \eqref{massivealpha}
	 	\be \alpha(\vec{q})=\left((c_\rho-d_\rho)a_{\vec{q}}+(c^*_\rho+d^*_\rho)a^\dagger_{-\vec{q}}\right)\left(2k\right)^{-\nu}~~\text{where}~~\Delta_-=\frac{d}{2}-\nu.\ee
	 	For this case
	 	\be |\alpha(\vec{q})\rangle=\left(\frac{q}{2}\right)^{-\nu}\left(c^*_\rho+d^*_\rho\right)|-\vec{q}\rangle,\ee
	 	so that
	 	\be  \langle \alpha(\vec{q})|\alpha(\vec{q})\rangle=|c_\rho+d_\rho|^2 \langle -\vec{q}|-\vec{q}\rangle, \ee  
	 	leading to the inner product 
	 	\begin{align}
	 \nn \frac{1}{\Omega}	\left(\alpha,\alpha\right)&= \frac{1}{\Omega}  \int\frac{d^{2h}q}{(2\pi)^{2h}}\langle \alpha(\vec{q})|\alpha(\vec{q})\rangle\\
	 	&=|c_\rho+d_\rho|^2
	 	\end{align}
which is once again  positive definite.
	 
	 \section{The connection between the Intertwining operator and the Shadow Transformation}
	 \label{sec:connection between intertwiner and shadow tr}	

The intertwining operator $G^+_{\chi}$ is crucial for the unitarity of the complementary series representations. We have seen its definition and how it acts in momentum space above with some examples. We have mentioned that the action of the interwining operator on a representation gives the shadow transformation. Here we would like to explicitly demonstrate that this is so.

\subsection{The relationship between the scaling dimensions}
The effect of shadow transformation is that it transforms an operator with scaling dimension $\Delta$ into an operator with scaling dimension $\tilde{\Delta}$ where the old and the new dimensions are related as
\be \label{eq:shadowtrdefprop} \Delta+\tilde{\Delta}=2h,\ee
in $2h$ spatial dimensions. The dimension $\Delta$ here is the scaling dimension of an operator as written in position space
\be \mathcal{O}_{\Delta}(\vec{x})\sim |\vec{x}|^{-\Delta},\ee
yet so far we have the operators explicitly written in momentum space. As summarized in table \ref{summary of the operators}, schematically our operators in momentum space have the form

\be \label{eqn:generalformforO}\mathcal{O}_\Delta(\vec{q})=q^c\left[Aa_{\vec{q}}+ B a^\dagger_{-\vec{q}}\right]\ee
 where $A,B$ are some coefficients and $c$ is a complex number. For the complementary series representations $c$ is real and can be positive or negative. Let us first confirm that this form indeed gives the scaling dimension $\Delta=h+c$ for $\mathcal{O}_{\Delta}(\vec{x})$. The position and momentum space operators are related to each other by a Fourier transform
 \be \label{eqn:fouriertr1} \mathcal{O}_\Delta(\vec{x})=\int \frac{d^{2h}q}{(2\pi)^{2h}}\mathcal{O}_\Delta(\vec{q})e^{i\vec{q}\cdot\vec{x}}.\ee
 It is not so easy to perform this integration but it is easy to read off the scaling dimension. As
 \begin{align}
 \nn \vec{x}&\to \lambda \vec{x},\\
 \vec{q}&\to \frac{q}{\lambda}
 \end{align} 
 using the commutation relation $[a_{\vec{q}},a^\dagger_{\vec{q}'}]=(2\pi)^{2h}\delta^{(2h)}(\vec{q}-\vec{q}')$ and the scaling property $\delta{^{(2h)}}(\lambda \vec{q})=\frac{\delta(\vec{q})}{|\lambda|^{2h}}$, it can be seen that the ladder operators scale as
 \begin{align}
\nn a_{\lambda^{-1}\vec{q}}&=\lambda^h a_{\vec{q}}\\
a^\dagger_{\lambda^{-1}\vec{q}}&=\lambda^ha^\dagger_{\vec{q}}.
 \end{align}
  With these scaling relations, the form \eqref{eqn:generalformforO} scales as
  \be \label{eq:scalingofOinmomentumsp}\mathcal{O}_{\Delta}(\lambda^{-1}\vec{q})=\left(\lambda^{-1}\right)^{c-h}\mathcal{O}_{\Delta}(\vec{q}).\ee
  To obtain the scaling dimension for $\mathcal{O}_{\Delta}(\vec{x})$, let's rescale \eqref{eqn:fouriertr1}
 \begin{align}\label{rescalefouriertr1} \nn\mathcal{O}_\Delta(\lambda\vec{x})&=\int \lambda^{-2h} \frac{d^{2h}q}{(2\pi)^{2h}}\mathcal{O}_\Delta(\lambda^{-1}\vec{q})e^{i\frac{\vec{q}}{\lambda}\cdot\lambda\vec{x}}\\
\nn &=\int \lambda^{-2h} \frac{d^{2h}q}{(2\pi)^{2h}}\lambda^{h-c}\mathcal{O}_\Delta(\vec{q})e^{i\vec{q}\cdot\vec{x}}\\
\nn&=\lambda^{-(h+c)}\int\frac{d^{2h}q}{(2\pi)^{2h}}\mathcal{O}_\Delta(\vec{q})e^{i\vec{q}\cdot\vec{x}}\\
 \mathcal{O}_\Delta(\lambda\vec{x})&=\lambda^{-(h+c)}\mathcal{O}_{\Delta}(\vec{x}).
 \end{align} 
 Thus indeed the operators of table \eqref{summary of the operators} written in $q$-space correspond to operators with scaling dimensions $\Delta=h+c$ in $x-$space.
 
 \subsection{Shadow Transformation in momentum space}
 Using the expression for the intertwining operator in momentum space we will now obtain the scaling dimension of the intertwined operator $\tilde{\mathcal{O}}_{\tilde{\Delta}}=G_{\chi}\mathcal{O}_\Delta(\vec{q})$ and check if this indeed satisfies the requirement \eqref{eq:shadowtrdefprop} for a shadow transformation. 
 
 Firstly, remember the subtlety in the definition of the intertwining operators, $G_\chi$ and $G_{\tilde{\chi}}$
 \begin{align}
 G_\chi&:C_{\tilde{\chi}}\to C_\chi,~~\text{for}~~c<0,\\
 G_{\tilde{\chi}}&: C_{\chi}\to C_{\tilde{\chi}},~~\text{for}~~c>0.
 \end{align}
 Either $G_\chi$ or $G_{\tilde{\chi}}$ is well defined for a given representation with weight $c$.
 The complementary series operators $\alpha(\vec{q})$ with $\Delta_\alpha=h-\nu$ and $\beta_\Delta(\vec{q})$ with $\Delta_\beta=h+\nu$ where $\nu\equiv|\sqrt{h^2-\frac{m^2}{H^2}}|$ belong to $C_{\chi}$ and their intertwined versions $\tilde{\alpha}(\vec{q})$, $\tilde{\beta}(\vec{q})$ with $\tilde{\Delta}_\alpha$ and $\tilde{\Delta}_\beta$ respectively belong to $C_{\tilde{\chi}}$. Since the operators $\alpha(\vec{q})$ have $c=-\nu<0$, the well defined intertwining operator for them is $G_\chi$. That means the well defined transformation is 
 \be \alpha(q)=G_{\chi}\tilde{\alpha}(\vec{q})~~\text{with}~~G^+_{\chi=\{0,c\}}(q)=\left(\frac{q^2}{2}\right)^c=\left(\frac{q^2}{2}\right)^{-\nu}.\ee
 On the other hand the operators $\beta(\vec{q})$ have $c=\nu>0$ and thus the well defined intertwining operator for them is $G_{\chi}$ leading to the transformation
 \be \tilde{\beta}(\vec{q})=G_{\tilde{\chi}}(q)\beta(\vec{q})~~\text{with}~~G^+_{\tilde{\chi}=\{0,\tilde{c}\}}(q)=\left(\frac{q^2}{2}\right)^{\tilde{c}=-c}=\left(\frac{q^2}{2}\right)^{-\nu}.\ee	
 We will study these two transformations individually to first obtain $\tilde{\mathcal{O}}_{\tilde{\Delta}}(\vec{q})$ and then read off the scaling dimension by scaling the position space operator 
 \be \label{eqn:tildeopinpositionspace}
 \tilde{\mathcal{O}}_{\tilde{\Delta}}(\vec{x})=\int \frac{d^{2h}q}{(2\pi)^{2h}}\tilde{\mathcal{O}}_{\tilde{\Delta}}(\vec{q})e^{i\vec{q}\cdot\vec{x}}.\ee
	 	
For the operators $\alpha(\vec{q})=q^{-\nu}\left[Aa_{\vec{q}}+Ba^\dagger_{-\vec{q}}\right]$ with $\Delta_\alpha=h-\nu$, the transformation
\begin{align}
\nn \alpha(q)&=G_{\chi}\tilde{\alpha}(\vec{q})\\
q^{-\nu}\left[Aa_{\vec{q}}+Ba^\dagger_{-\vec{q}}\right]&=\left(\frac{q^2}{2}\right)^{-\nu}\tilde{\alpha}(\vec{q})
\end{align}	
can be used to read off the corresponding operator in $C_{\tilde{\chi}}$, which is
\be \label{eqn:tildealpha}\tilde{\alpha}(\vec{q})=\left(\frac{q}{2}\right)^\nu\left[Aa_{\vec{q}}+Ba^\dagger_{-\vec{q}}\right].\ee
This operator scales as 
\be \tilde{\alpha}(\lambda^{-1}\vec{q})=\lambda^{h-\nu}\tilde{\alpha}(\vec{q}).\ee
In position space, 
\begin{align}
\tilde{\alpha}(\lambda\vec{x})=\lambda^{-(h+\nu)}\tilde{\alpha}(\vec{x}).
\end{align}
This implies that the scaling dimension for the intertwined operator $\tilde{\alpha}(\vec{x})$ is $\tilde{\Delta}_\alpha=h+\nu$.

The scaling dimension $\Delta_{\alpha}=h-\nu$ of $\alpha(\vec{x})$ and the  scaling dimension after intertwining, $\tilde{\Delta}_\alpha$ of $\tilde{\alpha}(\vec{x})=h+\nu$, satisfy the relation $\Delta+\tilde{\Delta}=2h$. Therefore the operators $\alpha(\vec{x})$ and $\tilde{\alpha}(\vec{x})$ related to each other by an intertwining operato are shadow transformations of each other. 

For the operators $\beta(\vec{q})= q^\nu \left[Aa_{\vec{q}}+Ba^\dagger_{-\vec{q}}\right]$ with $\Delta_\beta=h+\nu$ it is more straight forward to obtain the intertwined operator
\begin{align}
\nn \tilde{\beta}(\vec{q})&=G_{\tilde{\chi}}(q)\beta(\vec{q})~~\text{with}~~G^+_{\tilde{\chi}=\{0,\tilde{\nu}\}}(q)=\left(\frac{q^2}{2}\right)^{-\nu}\\
\tilde{\beta}(\vec{q})&=(2q)^{-\nu}\left[Aa_{\vec{q}}+Ba^\dagger_{-\vec{q}}\right],
\end{align}
which rescales as $\tilde{\beta}(\lambda^{-1}\vec{q})=\lambda^{h+\nu}\tilde{\beta}(\vec{q})$. In position space this scaling implies
\be \tilde{\beta}(\lambda\vec{x})= \lambda^{-(h-\nu)}\tilde{\beta}(\vec{x}).\ee
Thus for the operator $\beta(\vec{x})$ with $\Delta_{\beta}=h+\nu$, its intertwined version $\tilde{\beta}(\vec{x})$ has the scaling dimension $\tilde{\Delta}_\beta=h-\nu$. Again $\Delta_\beta+\tilde{\Delta}_\beta=2h$ and thus $\beta$ and $\tilde{\beta}$ are shadow transforms of each other. Hence the intertwining operation works as the shadow transformation.
 
\section{Conclusions and Outlook}
\label{sec:conclusion}
In this section we give a summary of our results and then point out possible considerations which we leave for future work. 
We end with a discussion comparing the notion of unitarity we have considered here for fields on de Sitter 
and the notion of (non)-unitarity in the dS/CFT literature.	 	
	
	\subsection{Summary of Results} 	
In this work we identified the late time behavior of scalar fields on de Sitter in arbitrary dimensions in terms of the unitary irreducible representations 
of the de Sitter symmetry group.
One of the results in the representation theory of groups is that, symmetries are represented either by \emph{unitary and linear} operators \cite{Wigner1902} 
or by \emph{anti-unitary and anti-linear} operators \cite{weinberg_1995}. Here, by studying the principal and complementary series 
representations of the de Sitter group $SO(2h+1, 1)$, we have discussed that light and heavy scalar fields at the late time boundary of 
de Sitter with Bunch-Davies initial conditions are all realized by unitary representations. 
The boundary operators, identified throughout the text for different mass ranges,  are summarized in table \ref{summary of the operators}.
	 	  
\begin{center}
\begin{table}[h]
\begin{tabular}{|c|c|c|c|}
\hline
& Category & Boundary operators $\Ocal_\Delta$ & $\Delta$\\
\hline
& & $Re(\nu)>0$ &\\
& &$\alpha^{I}(\vec{q})= -\frac{i}{\pi} \Gamma(\nu)\left(a_{\vec{q}}-a^\dagger_{-\vec{q}}\right)\left(\frac{q}{2}\right)^{-\nu},$ & $\Delta^{I}_\alpha=h-\nu$
  \\ 
Complementary & &$\beta^{I}(\vec{q})=\frac{1}{\Gamma(\nu+1)}\left(a_{\vec{q}}+a^\dagger_{-\vec{q}}\right)\left(\frac{q}{2}\right)^{\nu},$ & $\Delta^{I}_\beta=h+\nu$
 \\
&  $\frac{m^2}{H^2}<h^2$  & &\\
\cline{3-4}
Series & & $\nu=-\frac{2j+1}{2}$ where $j=0,1,...$ & 
\\
&&$\alpha^{II}(\vec{q})=\frac{1}{\Gamma(\frac{1-2j}{2})}\left(a_{\vec{q}}+a^\dagger_{-\vec{q}}\right)\left(\frac{q}{2}\right)^{-\frac{2j+1}{2}},$ & $\Delta^{II}_\alpha=h-\frac{2j+1}{2}$\\
& &$\beta^{II}(\vec{q})= -\frac{i}{\pi}\Gamma(-\frac{2j+1}{2})\left(a_{\vec{q}}-a^\dagger_{-\vec{q}}\right)\left(\frac{q}{2}\right)^{\frac{2j+1}{2}}$,  & $\Delta^{II}_\beta=h+\frac{2j+1}{2}$\\
& & $\alpha^{II}(\vec{q})\sim \tilde{\beta}^{I}$,   $\alpha^{I}(q)\sim \tilde{\beta}^{II}(q)$ & \\	
\hline
& $\frac{m^2}{H^2}=h^2$& $ \alpha^{\nu=0}(q)= a_{\vec{q}}+a^\dagger_{-\vec{q}}$ &$\Delta^{\nu=0}=h$ \\
Principal& $\nu=0$& & \\
\cline{2-4}
Series & $\frac{m^2}{H^2} > h^2,$ & $\alpha(\vec{q})=\left[(c_\rho-d_\rho)a_{\vec{q}}+(c^*_\rho+d^*_\rho)a^\dagger_{-\vec{q}}\right]\left(\frac{q}{2}\right)^{-i\rho},$ & $\Delta_-=h-i\rho$\\
& $\nu=i\rho$ &$\beta(\vec{q})=\left[(c_\rho+d_\rho) a_{\vec{q}}+(c^*_\rho-d^*_\rho)a^\dagger_{-\vec{q}}\right]\left(\frac{q}{2}\right)^{i\rho},$ & $\Delta_+=h+i\rho$
 \\
  & & &\\
\hline
\end{tabular}
\caption{Late time boundary operators. We have categorized the solutions among Complementary Series into two branches, and these branches are related to each other by a shadow transformation, denoted by a tilde. In the table $\nu^2=h^2-\frac{m^2}{H^2}$ where by $\nu$ and $\rho$ we denote the positive root. The scaling weight for the operators on the de Sitter late time boundary are: for $\alpha$,  $c_{\alpha}=-\nu$ and $Re(c_{\alpha})<0$ while for $\beta$, $c_{\beta}=+\nu$ and $Re(c_{\beta})>0$.}
\label{summary of the operators}
\end{table}
\end{center}

The operators in table \ref{summary of the operators} belong to the Hilbert space which is composed of the function space $C_\chi$, that is, 
the space of functions that obey the covariance condition \eqref{covariancecond}, and is equipped with the $K$-invariant scalar product,
 reviewed in section \ref{sec:elementsofG}. Most of our efforts were in investigating the properties of this 
scalar product throughout sections \ref{sec:unitaryreps} and \ref{sec:positivenormandunitarity}. The subtle difference 
between the two categories was that the well defined scalar product was straightforward 
for  Principal Series representations but  involved an intertwining operator for Complementary Series representations.

As introduced in section \ref{subsec:real weight c}, 
the finite scalar product for Complementary Series representations is defined as, $(\alpha,G^+_{\{\ell,\tilde{c}\}}\alpha)$ where $G^+_{\{\ell,\tilde{c}\}}$ 
is called the intertwining operator. We demonstrated its use in section \ref{subsec:conformally coupled scalar}. Moreover,  the action of 
the intertwining operator was recognized as a shadow transformation in section \ref{sec:connection between intertwiner and shadow tr}, 
by considering its effect on the scaling dimension. For reference, conformally coupled scalars belong to this category with $c=\pm\frac{1}{2}$. 

The scaling weight $c$ carries the information about the coordinate momentum dependence of the operator. Accordingly $c$ determines the $\vec{q}$-dependence 
of the intertwining operator as well. For the case of $c=0$ the inner product doesn't involve any extra q-dependence. Therefore no intertwining operators are necessary for this case, or in other words the role of the intertwining operator is reduced to an overall constant, and hence this operator belongs to the Principal Series representations.

\subsection{Considerations for future work}

The principal and complementary series representations we discussed here are constructed with respect to the homogeneous space $G/NAM$. It is important to stress that the principal and complementary series representations are not highest weight representations.\footnote{The subgroup $K=SO(2h+1)$ is a compact Lie group and finite dimensional irreducible representations of a compact lie group are completely characterized by their highest weight, $\ell$, which is the largest eigenvalue of the Lie algebra element that generates rotations along the z-axis. Out of the irreducible representations which are induced by the compact subgroup, the ones that have the highest weight $\underline{\ell}=(0,....,0,\ell)$ are referred to as \emph{type I representations}. Moreover, there is a connection between the representations of $M=SO(2h)$, which concern us more, and the type I representations of the compact subgroup $K=SO(2h+1)$. The connection is that type I representations of $SO(2h+1)$ are only decomposed into type I representations of $SO(2h)$.} It is possible to create other homogeneous spaces by taking 
the quotient with different subgroups. In addition to the representations we have discussed, the group $SO(2h+1,1)$ contains a third class of irreducible unitary representations, 
called the \emph{discrete series representations}, which are constructed with respect to, for example, the homogeneous space $G/K$ \cite{doi:10.1063/1.1704835}. 
The discrete series representations are an example of highest weight representations. Another category of highest weight representations are
 the \emph{exceptional series representations}. 
We leave the study of these later two classes for a  future work. 

Based on our analysis, one would expect that massless scalars with $c=\pm h$ to be a part of Complementary Series representations. 
However, the authors of \cite{Basile:2016aen} identify massless scalars as part of the exceptional series and so we also leave 
the categorization of this case for future work.

We note some possible generalizations of the work presented here.  For instance, the results here can be generalized to 
other spins by considering the appropriate spin dependent coefficient $K_{\ell s}(c)$ 
and projection operator $\Pi^{\ell s}(q)$ defined by \eqref{P_ell_s_def} in section \ref{subsec:real weight c}. 
Another possibility involves interactions more intricate than a simple mass term. Then the mode functions would involve something 
other than the Hankel functions and in that case, one would need to consider the late time limit of the corresponding 
function, leading to different operators than the ones in table \ref{summary of the operators}.

Recently, the representations of $SO(2h+1,1)$ have gained attention in the context of the dS/CFT proposal. This proposal was initially introduced by studying the early time boundary of de Sitter \cite{Strominger:2001pn} and considerations of the construction of a quantum Hilbert space for asymptotically de Sitter spacetimes \cite{Witten:2001kn}. Prior to the dS/CFT proposal, there exist calculations of entropy for asymptotically $dS_3$ spacetimes \cite{Maldacena:1998ih,Park:1998qk,Banados:1998tb} which demonstrate that the $dS_3$ entropy corresponds to that of a CFT. These works also identify a corresponding Virasoro algebra and specify its classical central charge by considering the asymptotic algebra for diffeomorphism invariance in \cite{Park:1998qk} and by making use of the Sugawara construction in \cite{Maldacena:1998ih,Banados:1998tb}. In all these works the key point relies on the fact that Einstein gravity in three dimensions with a positive cosmological constant is equivalent to Chern-Simons theory. Along this line, in \cite{Park:1998yw} a $Chern-Simons_{2+1}/CFT_1$ correspondance is obtained by showing that the Dirac algebra of the Noether charge for gauge invariance of Chern-Simons theory corresponds to a Kac-Moody algebra and the Dirac algebra of the Noether charge for diffeomorphism invariance of Chern-Simons theory corresponds to a Virasoro algebra with central charge. Since then, efforts of describing more concrete realizations of the dS/CFT \cite{Anninos:2011ui,Ng:2012xp,Anninos:2013rza} led to the recent work of \cite{Anninos:2017eib}, where Higher Spin Fields on de Sitter were associated to a bosonic $O(2\Ncal)$ vector model with $2\Ncal$ fields $Q$, at the late time boundary of de Sitter
 and the Hilbert Space and the partition function for correlators of this model were constructed.
 The Higher Spin theory on de Sitter only accommodates conformally coupled fields of section \ref{subsection:special mass soln}, among its scalar sector. From a low energy effective field theory point of view, one should be able to consider all cases mentioned in section \ref{sec:ltandboundaryops} as all these fields are allowed by 
the symmetries of de Sitter. This raises the question: can there be other realizations of the dS/CFT proposal, 
 or is there some condition which forbids the scalars in the other cases of section \ref{sec:ltandboundaryops}?		
	 		
Shortly after the dS/CFT proposal, it was noticed that this formalism could be developed into a new way of calculating inflationary correlators, 
by including deformation operators to the CFT \cite{Strominger:2001gp,Maldacena:2002vr,vanderSchaar:2003sz}. 
We hope to address such issues in a future work.
%

\subsection{Notions of unitarity on dS and in dS/CFT literature}
The operators listed in table \ref{summary of the operators} are the boundary operators from a bulk perspective, as they were obtained by considering the late time limit of solutions in de Sitter. The connection between these boundary operators and the CFT operators is not that they are the same but that they are expected to be related to each other in terms of their scaling dimensions and correlation functions. 

The boundary operators from the bulk de Sitter solutions have positive definite norm and are therefore unitary as we have discussed. The unitarity of the CFT operators is a different 
matter altogether\footnote{We thank Dionysios Anninos for clarifying this point for us.}. Unitarity defined as states having a positive definite norm applies both to Quantum Field Theory on de Sitter and to Conformal Field Theories. One subtle difference is that in the case of a Conformal Field Theory the positive definite norm restricts scaling dimensions to be real.
Comparing dS/CFT with AdS/CFT, the boundary in the case of AdS is at spatial infinity along one of the spatial directions, 
therefore it is a Lorentzian surface and the corresponding CFT is Lorentzian.
In the case of dS, the boundary is reached through the late time limit, and thus, it is a Euclidean surface as well as the corresponding CFT.
 While unitarity is a crucial property for Lorentzian CFT's, it is not a necessary condition for Euclidean CFT's \cite{Blumenhagen:2009zz}. 
	
The earlier example of dS/CFT, the Euclidean $Sp(\Ncal)$ CFT initially proposed as the dual to Vasiliev Higher Spin gravity on $dS_4$ 
was noted to be non-unitary~\cite{Anninos:2011ui,Ng:2012xp}. 
An explicit calculation in \cite{Ng:2012xp} shows that the CFT operators have negative norm.
 The more recent  bosonic $O(2\Ncal)$ vector model,
referred to as the Q-model and proposed as a microscopic description for Vasiliev gravity on de Sitter,
 is noted to have a Hilbert space of positive definite operators~\cite{Anninos:2017eib}. 
The single trace primaries in that proposal correspond to the shadows 
of the higher weight de Sitter boundary operators, $\tilde{\beta}_I$ where $I$
 here involves the spin label as well. The authors point out, however, that 
it seems hard to reconstruct the other boundary operators $\alpha_I$ and that normalizable perturbative single particle states 
on the bulk QFT do not correspond to normalizable states if they are to be interpreted as states on the Q-model Hilbert Space for spin higher than zero.

 A detailed study on the relation between the bulk correlation functions under the late time limit to the CFT correlation functions have been presented in \cite{Maldacena:2002vr,Anninos:2014lwa} via calculating the wavefunction in various cases.
 These discussions show that the late time limit of the wavefunction can be taken as an approximation of the CFT partition function from which the CFT correlators can be computed. Moreover both works show that the wavefunction of de Sitter and Euclidean Anti de Sitter, and hence the correlation functions, are related to each other by an analytic continuation. 
Here we studied the classification of the boundary fields obtained from the bulk fields on de Sitter, on their own right. This can be seen as a complementary approach to transferring results from Anti de Sitter to de Sitter via analytic continuation.    
 We hope this study will serve as a first step towards a group theoretical approach of fields of general 
spin on de Sitter which may eventually add to the existing frameworks for studying perturbations during epochs of inflation or dark energy domination.

\acknowledgments
We are greatful to Dionysios Anninos for numerous discussions and his encouragement of this work. We deeply thank Joris Raeymaekers for sharing his insight on higher spin fields and holography with us through valuable discussions and for his helpful comments on an earlier version of this manuscript. We thank Jan Pieter van der Schaar, Metin Ar{\i}k, Guilherme L. Pimentel for general discussions about de Sitter spacetimes. We thank Mu-In Park, Lasse Schmieding and an anonymous referee for their comments on the previous version. GŞ was supported by IOP Researchers Mobility Grant $CZ.02.2.69/0.0/0.0/16\_027/0008215$ in the first half of this project and is supported by the European Union's Horizon 2020 research and innovation programme under the Marie Sk\l{}odowska-Curie grant agreement No 840709 since the second half.
CS is acknowledges support from the European Structural and Investment  Funds  and  the  Czech  Ministry  of  Education,  Youth  and  Sports  (MSMT)  
(Project  CoGraDS  -CZ.02.1.01/0.0/0.0/15003/0000437).
	 	\appendix
	 	\section{Further properties of the group $SO(2h+1,1)$}
	 	\label{app:furtherpropertiesofG}
	 	In section \eqref{sec:elementsofG}, following \cite{Dobrev:1977qv} we summarized the notation for group elements of $SO(2h+1,1)$, and the properties of its irreducible representations. Here we list some further properties of $SO(2h+1,1)$, such as how the generators of its Lie algebra can be realized, common decomposition of its group elements $g$ in terms of the elements of the subgroups and the irreducibility of its elementary representations. For further details and more properties we refer the reader to \cite{Dobrev:1977qv}. 
	 	\subsection{The defining properties of group $SO(\Ncal,1)$}
	 	
	 	As was mentioned in section \eqref{sec:elementsofG}, the group $SO(\Ncal,1)$, also denoted as $SO(2h+1,1)$, is the set of all linear transformations on the real $\Ncal+1$-dimensional vector space, which leave the quadratic form,
	 	\be \xi^2=\xi\eta\xi={\xi_1}^2+...+{\xi_{\Ncal}}^2-{\xi_0}^2~~\text{where}~~\eta_{\mu\nu}=diag(-1,1,...,1),\ee
	 	invariant. Its elements are $(\Ncal+1)\times(\Ncal+1)$ matrices, $g$ who satisfy the properties
	 	\be 
  g^{T}\eta g=\eta,~~\det g=1,~~{g^0}_0\geq1. 
\ee 
	 	
	 	The generators of the Lie algebra $\mfg$ of group $G$ are $(\Ncal+1)\times(\Ncal+1)$ matrices which satisfy
	 	\be X^{T}\eta+\eta X=0.\ee
	 	In the basis where $X_{\mu\nu}=-X_{\nu\mu}$ with $\mu,\nu=0,1,...,\Ncal$, they satisfy the  commutation relations
	 	\be\label{mathcommutation} [X_{\mu\nu},X_{\alpha\beta}]=\left(\eta_{\mu\alpha}X_{\nu\beta}+\eta_{\nu\beta}X_{\mu\alpha}\right)
   -\left(\eta_{\mu\beta}X_{\nu\alpha}+\eta_{\nu\alpha}X_{\mu\beta}\right),\ee
	 	and can be realized in matrix form via
	 	\be {\left(X_{\mu\nu}\right)^\alpha}_\beta=\eta_{\mu\beta}\delta^\alpha_\nu-\eta_{\nu\beta}\delta^{\alpha}_\mu.\ee
	 	This form of the generators differs from the form of physical generators by a factor of i, $J_{\mu\nu}=iX_{\mu\nu}$.
	 	
	 	\subsection{Killing vectors of $dS_4$}
	 	\label{sec:killing vectors}
 Here $h=3/2$ so that Greek indices take values in $0\ldots 3$ and lower case indices in $1\ldots 3$.
	 	We start our calculation in planar coordinates with conformal time, where the metric takes the form
	 	\be ds^2=\frac{1}{H^2\eta^2}\left[-d\eta^2+d\vec{x}^2\right].\ee 
	 	In these coordinates the nontrivial Christoffel symbols are ${\Gamma^\eta}_{\eta\eta}={\Gamma^\eta}_{ii}={\Gamma^i}_{i\eta}=-\frac{1}{\eta}$. Making use of the metric compatibility $\nabla_\mu g_{\nu\kappa}=0$, we can write the Killing equation $\nabla_\mu\xi_\nu+\nabla_\nu\xi_\mu=0$ as
	 	\be g_{\nu\alpha}\nabla_\mu\xi^\alpha+g_{\mu\alpha}\nabla_{\nu}\xi^\alpha=0.\ee
	 	The solution to this equation gives 10 Killing vectors, which is the number of $S0(4,1)$ generators.	 	 These are
\begin{itemize}
	 	\item One Killing vector  generating dilatations parametrized by $\lambda$
\be 
  \label{xidilatation} 
  \overset{{\rm (D)}}{\xi}=\lambda\eta\partial_\eta+\lambda x^i \partial_i.
 \ee
\item Three Killing vectors generating translations parametrized by the vector $\vec{a}$
\be 
 \label{xitr} 
\overset{{\rm (Trans)}}{\xi}=a^i\partial_i
\ee
\item Three Killing vectors generating Special Conformal Transformations parametrized by the vector $\vec{b}$
\be 
\label{xiSCT} 
\overset{{\rm (SCT)}}{\xi}=-2\eta\left(\vec{x}\cdot\vec{b}\right)\partial_\eta
  + \left[|\vec{x}|^2b^i-2x^i\left(\vec{x}\cdot\vec{b}\right)-\eta^2b^i\right]\partial_i.
\ee
\item Three Killing vectors associated with spatial rotations, parametrized by a $3\times 3$ antisymmetric tensor  $\omega$
	 	\be 
\label{xirot} 
\overset{{\rm (Rot)}}{\xi}=\frac{1}{2}\left({\omega^i}_jx^j\partial_i-{\omega_j}^ix^j\partial_i\right).
\ee
\end{itemize}
	 	
Transforming to  planar coordinates in terms of cosmological time via $\eta\to t=-\frac{\ln(-H\eta)}{H}$ then
\be
 ds^2=-dt^2+e^{2Ht}d\vec{x}^2
\ee
	 	we find that the Killing vector associated with dilatation becomes
	 	\be \label{xidil_t} \overset{(D)}{\tilde{\xi}}=-\frac{\lambda}{H_{dS}}\partial_t +\lambda x^i\partial_i;~i=1,2,3.\ee
	 	Note that by equation \eqref{xitr} there are no Killing vectors that generate time translation, hence time translations are not a symmetry of de Sitter.\footnote{For
 completeness, de Sitter in terms of static patch coordinates where, $ds^2=-(1-H^2r^2)d\tau^2+\frac{dr^2}{1-H^2r^2}+r^2d\Omega^2$,
 does have a timelike Killing vector. In fact the static patch coordinates were invented in the pursuit of finding a coordinate system for de Sitter 
which will have a timelike Killing vector. For example in terms of Static Patch coordinates one can define energy eigenstates of the Hamiltonian, where 
as one cannot do so in terms of planar coordinates. However, the static patch does not cover the entire de Sitter manifold and as such the timelike Killing vector is not
globally defined. } 
In terms of cosmic time however, the dilatation symmetry, equation \eqref{xidil_t}, contains something like time translation.
	 	
	 	Above we have written the Killing vectors in a way where the parameters associated with the transformations they generate are explicit. The appropriately normalized generators are \cite{Baumann:2019oyu}
	 	\begin{subequations}
	 	\begin{align}
	 	\text{Dilatations:}~& D=-\eta\partial_\eta-x^i\partial_i\\
	 	\text{Translations:}~& T_i=\partial_i\\
	 	\text{SCT:}~& C_i=2x_i\eta\partial_\eta+\left[2x^jx_i + \left(\eta^2-|\vec{x}|^2\right)\delta^j_i\right]\partial_j\\
	 	\text{Rotations:}~& X_{ij}=x_i\partial_j-x_j\partial_i.
	 	\end{align}
	 	\end{subequations}
	 	Normalized as such, these generators satisfy the commutation relations \eqref{mathcommutation}.

	 	\subsection{Decomposition of the group elements $g$ in terms of the elements of the subgroups}
	 	\label{sec:decompositionofg}
	 	The group G acts in a natural way, that is by left translation, on the homogeneous space $G/{NAM}\sim K/M\sim S^{2h}$. However, it is necessary to be able to note how the group G acts on Euclidean space $\RR^{2h}$ with elements $\vec{x}\in \RR^{2h}$ where the fields live. The vector space $\RR^{2h}$ can be identified 
with the right coset $\tilde{n}NMA$.\footnote{To clarify the notation, we reserved capital letters for the groups and lowercase letters for the group elements. When the subgroup $NAM$ is factored out of $G$ what practically remains is the subgroup of translations $\tilde{N}$. However you can define an equivalence class for the elements $\tilde{n}\in \tilde{N}$ and it is this equivalence class that the notation $\tilde{n}NMA$ aims to capture.} Thus there is a unique correspondence between the elements $x$ and the elements $\tilde{n}$, 
	 	\be x=\tilde{n}NMA\ee
	 	which is also denoted by $\tilde{n}_x$. 
	 	
	 	The connection of the group elements $g$ to the Euclidean space element $x$ is explicit in the so called Bruhat decomposition
	 	\be \text{Bruhat decomposition:}~~~g=\tilde{n}nam.\ee
	 	Notice that this decomposition explicitly carries information about SCT, and the Euclidean Lorentz group in addition to translations and dilatations. Another decomposition, the Iwasawa decomposition, on the other hand brings forth the compact nature of the group elements
	 	\be \text{Iwasawa decomposition:}~~~g=kna=\tilde{n}ak.\ee

	 	\section{The normalized  intertwining operator $G_\chi$}
	 	\label{app:normGX2}
	 	Here we will follow the derivation of \cite{Dobrev:1977qv} for obtaining the intertwining operator $G_\chi$ and highlight some of the in between steps. Remember that this operator is defined as
	 	\be G_\chi: C_{\tilde{\chi}} \to C_\chi,~~\text{and}~~G_\chi=\gamma_\chi A_\chi \Ical(I_s).\ee
	 	Here we consider the equivalence map such that $\tilde{l}=l$. By this definition,
	 	\be \left[G_\chi\tilde{\mff}\right](\tilde{n}_{\vec{x}_1})=\gamma_\chi A_\chi\left[\Ical(I_s)\tilde{\mff}\right](\tilde{n}_{\vec{x}_1}).\ee
	 	The action of the equivalence map $\Ical(I_s)$ on functions is
	 	\be \left[\Ical(I_s)\tilde{\mff}\right](\tilde{n}_{\vec{x}_1})=D^\ell(I_s)\tilde{\mff}(\tilde{n}_{\vec{x}_1}),\ee
	 	which leads to
	 	\be \left[G_\chi\tilde{\mff}\right](\tilde{n}_{\vec{x}_1})=\gamma_\chi D^\ell(I_s) \left[A_\chi\tilde{\mff}\right](\tilde{n}_{\vec{x}_1}).\ee
	 	As was noted in equation \eqref{int1}
	 	\be \left[A_\chi\tilde{\mff}\right](g)=\int_{\tilde{N}}\tilde{\mff}(gw\tilde{n}_{\vec{x}})d^{2h}x.\ee
	 	For $g=\tilde{n}_{\vec{x}_1}$ this means
	 	\be \left[G_\chi\tilde{\mff}\right](\tilde{n}_{\vec{x}_1})=\gamma_\chi D^\ell(I_s)\int_{\tilde{N}}\tilde{\mff}(\tilde{n}_{\vec{x}_1}w\tilde{n}_{\vec{x}})d^{2h}x.\ee
	 	Now we make use of the identity
	 	\be w\tilde{n}_{\vec{x}}=\tilde{n}_{\vec{x}'}n_{I_s\vec{x}}a(\vec{x},w)m(\vec{x},w),\ee
	 	with $\vec{x}'=I_sR\vec{x}$ where $R$ is the conformal inversion. This leads to
	 	\be \left[G_\chi\tilde{\mff}\right](\tilde{n}_{\vec{x}_1})=\gamma_\chi \int_{\tilde{N}}D^\ell(I_s)\tilde{\mff}(\tilde{n}_{\vec{x}_1}\tilde{n}_{\vec{x}'}n_{I_s\vec{x}} a m)d^{2h}x.\ee
	 	The elements of the translation group obey
	 	\be \tilde{n}_{\vec{x}_1}\tilde{n}_{\vec{x}'}=\tilde{n}_{\vec{x}_1+\vec{x}'},\ee
	 	and for $\tilde{\mff}\in \Ccal_{\tilde{\chi}}$ the covariance condition reads
	 	\be \tilde{\mff}(\tilde{n}_{\vec{x}_1+\vec{x}'}n_{I_s\vec{x}}am)=|a|^{h-c}D^\ell(m)^{-1}\tilde{\mff}(\tilde{n}_{\vec{x}_1+\vec{x}'}).\ee
	 	With the identity $\tilde{\mff}(\tilde{n}_{\vec{x}})=\tilde{f}(\tilde{n}_{\vec{x}})$ and $|a|=\frac{1}{x^2}$ we arrive at 
	 	\begin{align} \left[G_\chi\tilde{f}\right](\vec{x}_1)&=\gamma_\chi\int|a|^{h-c}D^\ell(I_s)D^\ell(m)^{-1}\tilde{f}(\vec{x}_1+\vec{x}')d^{2h}x\\
	 	\label{appGX}&=\gamma_\chi\int D^\ell(I_sm^{-1})\tilde{f}(\vec{x}_1+\vec{x}')(x^2)^{c-h}d^{2h}x.\end{align}
	 	Remember that so far $\vec{x}'=I_sR\vec{x}$. It is convenient to make a change of coordinates such that
	 	\be \vec{x} \to \vec{x}': \vec{x}=I_sR\vec{x}'~~\text{and}~~x^2=\frac{1}{x'^2},~~d^{2h}x=(x'^2)^{-2h}d^{2h}x'\ee
	 	 in order to put equation \eqref{appGX} in a more operational format, which reads
	 	 \be \left[G_\chi\tilde{f}\right](\vec{x}_1)=\gamma_\chi \int D^\ell(I_sm^{-1})\tilde{f}(\vec{x}_1+\vec{x}')(x'^2)^{-(h+c)}d^{2h}x'.\ee
	 	 Further more, to handle $D^\ell(I_sm^{-1})$ make the following transformation
	 	 \be \vec{x}'\to \vec{x}_2:\vec{x}'=\vec{x}_2-\vec{x}_1\ee
	 	 which gives
	 	 \be\label{apptr2} \left[G_\chi\tilde{f}\right](\vec{x}_1)=\gamma_\chi\int \frac{D^\ell(I_sm^{-1})}{|\vec{x}_2-\vec{x}_1|^{2(h+c)}}\tilde{f}(\vec{x}_2)d^{2h}x_2.\ee
	 	 By the following identities
	 	 \begin{subequations}
	 	 \begin{align}
	 	 m^{-1}&=r(\vec{x})I_s,\\
	 	 I_s r(I_s R\vec{x}')I_s&=r(\vec{x}')=r(\vec{x}_2-\vec{x}_1)
	 	 \end{align}
	 	 \end{subequations}
	 	 we have that
	 	 \be D^\ell(I_sm^{-1})=D^\ell(I_sr(\vec{x})I_s).\ee
	 	 Remember that $\vec{x}=I_sR\vec{x}'$ and so
	 	 \begin{align} D^\ell(I_sm^{-1})&=D^\ell(I_sr(\vec{x})I_s)\\
	 	 &=D^\ell(I_sr(I_sR\vec{x}')I_s)=D^\ell(r(\vec{x}'))=D^\ell(r(\vec{x}_2-\vec{x}_1)).
	 	 \end{align}
	 	 Eventually equation \eqref{apptr2} gives
	 	 \be \left[G_\chi\tilde{f}\right](\vec{x}_1)=\gamma_\chi \int \frac{D^\ell(r(\vec{x}_2-\vec{x}_1))}{|\vec{x}_2-\vec{x}_1|^{2(h+c)}}\tilde{f}(\vec{x}_2)d^{2h}x_2\ee
	 	as was promised in equation \eqref{intop}. 

 	\section{An Exemplary shadow transformation}
 	\label{app:exemplary shadow transforms}
 In section \ref{sec:light scalars in general} we studied the corresponding late time operators in two branches \eqref{branchrevg0boundary} and \eqref{-2n+1/2branch}, because our naming of the operators depended on whether $\nu$ is positive or negative. The first branch corresponded to $\Re(\nu)>0$ which includes $\nu=\frac{2j+1}{2}$, and the second branch was specified for $\nu=-\frac{2j+1}{2}$ for $j=0,1,2...$ Since both branches include operators with weight $\Delta=h\pm\frac{2j+1}{2}$, there should be some relation between the two branches. Here we demonstrate that the two operators are related to each other via shadow transformations.
 
 The operators of our focus for $\nu=\pm\frac{2j+1}{2}$ are as follows
 
 \begin{subequations}
 \label{branch1appendix}
 \begin{align}
\nn \text{{\bf Branch 1}}&\\
\label{branch1alphaapp}
 \alpha^{I}(\vec{q})&= -\frac{i}{\pi} \left(\frac{q}{2}\right)^{-\frac{2j+1}{2}}\Gamma(\frac{2j+1}{2})\left(a_{\vec{q}}-a^\dagger_{-\vec{q}}\right),~~\Delta^{I}_{\alpha}=h-\frac{2j+1}{2},~~c^{I}_{\alpha}=-\frac{2j+1}{2}\\
\label{branch1betaapp}\beta^{I}(\vec{q})
 &= \left(\frac{q}{2}\right)^{\frac{2j+1}{2}}\frac{1}{\Gamma(\frac{2j+3}{2})}\left(a_{\vec{q}}+a^\dagger_{\vec{-q}}\right),~~\Delta^{I}_{\beta}=h+\frac{2j+1}{2},~~c^{I}_{\beta}=\frac{2j+1}{2}
 \end{align}
 \end{subequations} 
 and
 \begin{subequations}
 \label{branch2appendix}
 \begin{align}
\nn \text{{\bf Branch 2}}&\\
\label{branch2alphaapp} 
\alpha^{II}(\vec{q})&=\left(\frac{q}{2}\right)^{-\frac{2j+1}{2}}\frac{1}{\Gamma(\frac{1-2j}{2})}\left(a_{\vec{q}}+a^\dagger_{-\vec{q}}\right),~~\Delta^{II}_{\alpha}=h-\frac{2j+1}{2},~~c^{II}_{\alpha}=-\frac{2j+1}{2}\\
\label{branch2betaapp}
\beta^{II}(\vec{q})&= -\frac{i}{\pi} \left(\frac{q}{2}\right)^{\frac{2j+1}{2}} \Gamma(-\frac{2j+1}{2})\left(a_{\vec{q}}-a^\dagger_{-\vec{q}}\right)~~\Delta^{II}_{\beta}=h+\frac{2j+1}{2},~~c^{II}_{\beta}=\frac{2j+1}{2}.
 \end{align}
 \end{subequations}	
 
 The operator $\beta^{II}$ has a positive weight $c_{\beta^{II}}=\frac{2j+1}{2}$. This means the well defined intertwining operator that can act on $\beta^{II}$ is $G^+_{[0,\tilde{c}_{\beta^{II}}]}(q)=\left(\frac{q^2}{2}\right)^{-\frac{2j+1}{2}}$. Acting on $\beta^{II}$ with $G^+_{[0,\tilde{c}_{\beta^{II}}]}(q)$ would give the shadow operator $\tilde{\beta}^{II}$.
 When worked out as follows
 \begin{align}
 \label{shadowofbetaII}
\nn \tilde{\beta}^{II}(\vec{q})&=G^+_{[0,\tilde{c}_{\beta^{II}}]}(q)\beta^{II}(\vec{q})\\
\nn &=
- \frac{i}{\pi} q^{-\frac{2j+1}{2}}\Gamma(-\frac{2j+1}{2})\left(a_{\vec{q}}-a^\dagger_{-\vec{q}}\right)\\
 &=\left(\frac{1}{2}\right)^{\frac{2j+1}{2}}\frac{\Gamma(-\frac{2j+1}{2})}{\Gamma(\frac{2j+1}{2})} \alpha^{I}(\vec{q})
 \end{align}
 this shadow operator turns out to be proportional to the operator $\alpha^{I}(q)$ of the first branch.
 
 Similarly the operator $\beta^{I}(\vec{q})$ has the positive weight $c_{\beta^{I}}=\frac{2j+1}{2}$, and acting on this operator with the well defined intertwining operator $G^+_{[0,\tilde{c}_{\beta^{I}}]}(q)=\left(\frac{q^2}{2}\right)^{-\frac{2j+1}{2}}$ gives the shadow operator
 \begin{align}
 \nn \tilde{\beta}^{I}(\vec{q})&=G^+_{[0,\tilde{c}_{\beta^{I}}]}(q)\beta^{I}(\vec{q})\\
 &=\left(\frac{1}{2}\right)^{\frac{2j+1}{2}}\frac{\Gamma(\frac{1-2j}{2})}{\Gamma(\frac{2j+3}{2})}\alpha^{II}(\vec{q}),
 \end{align} 
 which is proportional to the second branch operator $\alpha^{II}$. 
 
 Thus we can conlude that the branch one and branch two solutions addressed in section \ref{sec:light scalars in general} are related to each other via the shadow transformation. 

\bibliography{bibliography}
\bibliographystyle{JHEP}

\end{document}